\documentclass[12pt]{iopart}

\usepackage[T1]{fontenc}
\usepackage[latin9]{inputenc}

\usepackage{iopams}
\usepackage{setstack}

\usepackage{mathrsfs}
\usepackage{amssymb}
\usepackage{graphicx}
\usepackage{esint}
\usepackage{bm}
\usepackage{bbm}
\usepackage{bbold}
\usepackage[all]{xy}
\usepackage{enumitem}
\usepackage{hyperref}
\usepackage{color}
\usepackage[usenames,dvipsnames]{xcolor}

\def\cob{\color{blue}}
\def\cop{\color{Blue}}
\newcommand{\arX}[1]{\href{http://arxiv.org/abs/#1}{arXiv:{\cop #1}}}
\newcommand{\doin}[5]{\href{http://dx.doi.org/#1}{\cob \textit{#2} #3 \textbf{#4} #5}}
\newcommand{\doij}[6]{\href{http://dx.doi.org/#1}{\cob \textit{#2} #3 #4(#6)#5}}
\newcommand{\doit}[5]{\href{http://dx.doi.org/#1}{\cob #2 \textbf{#3} (#5) #4}}
\newcommand{\ndoin}[5]{\href{#1}{\cob  \textit{#2} #3 \textbf{#4} #5}}
\newcommand{\boxd}[1]{\boxed{\phantom{\Biggl(}#1\phantom{\Biggl)}}}
\newcommand{\Eq}[1]{(\ref{#1})}
\newcommand{\tia}[1]{#1}

\renewcommand\[{\begin{equation}}
\renewcommand\]{\end{equation}}

\newcommand{\ra}{\rightarrow}

\newcommand{\lolra}{\longleftrightarrow}

\newcommand{\In}{\subset}
\newcommand{\cd}{\cdot}

\newcommand{\R}{\mathbb R}

\newcommand{\sgn}{\mbox{sgn}}

\def\d{\mbox d}

\makeatother

\begin{document}

\begin{flushright} \doit{10.1088/0264-9381/30/12/125006}{Class.\ Quantum Grav.}{30}{125006}{2013} \hfill  \arX{1208.0354}\\ AEI-2012-078\end{flushright}

\title{Laplacians on discrete and quantum geometries}

\author{Gianluca Calcagni\footnote{Present address: Instituto de Estructura de la Materia --- CSIC,
calle Serrano 121, E-28006 Madrid, Spain.}, Daniele Oriti and Johannes Th\"urigen}
\address{Max Planck Institute for Gravitational Physics (Albert Einstein Institute),\\
Am M\"uhlenberg 1, D-14476 Potsdam, Germany}
\eads{\mailto{calcagni@iem.cfmac.csic.es}, \mailto{doriti@aei.mpg.de} and \mailto{johannes.thuerigen@aei.mpg.de}}

\begin{abstract}
We extend discrete calculus for arbitrary ($p$-form) fields on embedded lattices to abstract discrete geometries based on combinatorial complexes. We then provide a general definition of discrete Laplacian using both the primal cellular complex and its combinatorial dual. The precise implementation of geometric volume factors is not unique and, comparing the definition with a circumcentric and a barycentric dual, we argue that the latter is, in general, more appropriate because it induces a Laplacian with more desirable properties. We give the expression of the discrete Laplacian in several different sets of geometric variables, suitable for computations in different quantum gravity formalisms.
Furthermore, we investigate the possibility of transforming from position to momentum space for scalar fields, thus setting the stage for the calculation of heat kernel and spectral dimension in discrete quantum geometries.
\end{abstract}

\date{August 1, 2012}




\maketitle


\section{Introduction}
In a variety of current approaches to quantum gravity, including loop quantum gravity (LQG) \cite{Rovelli:2004wb,Thiemann:1111397} and spin-foam models \cite{Oriti:2001jh,Perez:2003wk,Perez:2012uz}, group field theory \cite{GFT1,GFT2}, simplicial quantum gravity, be it quantum Regge calculus \cite{Hamber:2009zz} or (causal) dynamical triangulations \cite{Ambjorn:2010kv}, the basic building blocks of geometry and spacetime are discrete in nature. Depending on the specific theory considered, these discrete building blocks can be interpreted as the true degrees of freedom of quantum spacetime (it is the case in loop gravity and spin foams, as well as in group field theory) or as a convenient regularization (in simplicial gravities) chosen only for the purpose of defining the theory or being able to calculate with it. In any case, one is left with the task of reconstructing a continuum spacetime and its geometry starting from such discrete structures (on which one can make appropriate superpositions of states, in canonical setting, or define histories, in path-integral-like frameworks). Despite a wealth of results obtained in recent years in all these approaches, the issue of recovering continuum geometry from discrete structures, or more generally that of extracting geometric information from them, remains outstanding. (A case where a continuum geometry arises naturally as a `blurring' of a discrete-symmetry structure is complex-order fractional spacetimes \cite{fra4}.) Notice also that the issue of determining the effective geometry in a given regime is present also in continuum frameworks like asymptotic safety \cite{Niedermaier:2006up}, simply because one allows for quantum fluctuations between continuum geometric configurations.

\

One difficulty has to do with the limited number of geometric observables being available and under control in all these scenarios at the quantum level, where it is clear that the only meaningful notion of effective geometry is in terms of the evaluation of specific quantum geometric observables. In LQG and spin foams, for example, one has good control over the definition of areas and 3-volumes as quantum operators and over their spectrum, and definitions of length and 4-volume observables exist, but do not come with a good enough analytic control. Various distance measures exist in the simplicial context, which are usually dealt with numerically. In general, it is fair to say that much more work is needed and that the more examples of geometric observables we can construct, the more the task of analyzing the effective geometries produced in our quantum gravity models will be facilitated.

More such observables could be defined in the case of quantum gravity coupled to matter, and matter is also expected to permit the construction of {\it local} geometric observables (as opposed to global ones) which are still diffeomorphism-invariant. Again, much on matter coupling in canonical and covariant approaches is known, even in the discrete context \cite{Thiemann:1998hn, Oriti:2002gy, Speziale:2007ha, Oriti:2006kg, Fairbairn:2007bu, Bianchi:2010vy, Han:2011uu, Rovelli:2010ic}, but this is another area where more results are needed.
One example of a geometric observable that has been widely used for `reconstruction purposes', i.e., as a probe of the geometry of states, phases or histories in quantum gravity models, is the spectral dimension \cite{AJL4,LaR5,Ben08,Mod08,Hor3,CaM,Car09,BeH,MPM,Car10,SVW1,SVW2,AA,frc4,RSnax}. Being defined from the trace of the heat kernel, it depends on the underlying geometry through the Laplacian operator and implicitly relies on some notion of matter field. 

\

In this paper, we focus on the notion of Laplacian in a discrete context. First of all, the Laplacian is an interesting geometric kinematical observable {per se}.  Second, it is the key ingredient for the definition of momentum space and, as we mentioned, of the spectral dimension. Third, it is needed to construct coupled gravity plus matter models, as it enters the propagators for matter fields (be them scalars or gauge fields). We set up a general, systematic approach to its construction, which can turn out to be useful for applications \cite{COT2}.

The plan is the following. In sections \ref{sec1} and \ref{sec2} we provide a coherent framework for the definition of functions, $p$-form fields and differential operators on fundamentally discrete (and, later, quantum) geometries, more specifically on abstract simplicial and cellular complexes. Although we base our systematic approach on the recently developed {\it discrete exterior calculus} of \cite{Desbrun:2005ug} (see also \cite{Hirani:2003ug,Desbrun:2003db}), we employ their explicit treatment of geometric volume factors to generalize to abstract complexes attached with dynamic geometrical variables. 

Using this, we propose (section \ref{sec2}) a general definition of the discrete Laplacian operator on arbitrary 
simplicial pseudo-manifolds. The definition makes use of both primal and dual complexes. We then study the properties of this Laplacian and, through them, compare different choices of geometry of dual complex (barycentric and circumcentric). 

We also describe (section \ref{sec3}) the generalization of the same operator to simplicial pseudo-manifolds with boundaries and arbitrary cellular complexes and the notion of momentum transform in terms of eigenfunctions of the Laplacian, that plays a crucial role in the calculation of the heat kernel trace, which we also discuss. 

We then show (section \ref{sec4}) the various expressions that the 
Laplacian operator takes in different choices of geometric variables, again having in mind the sets of variables currently used in various quantum gravity frameworks. This will facilitate concrete applications and computations \cite{COT2}. As already mentioned, the setting is chosen as general as possible. In particular, complexes
are defined only combinatorially in order to be applicable to diverse theories
of quantum geometry at a second stage. This is the type of complexes arising, for example, in group field theory \cite{GFT1, GFT2, Gurau:2010iu, Smerlak:2011ea, Gurau:2011hu}, spin-foam models and LQG  \cite{Oriti:2001jh,Perez:2003wk,Perez:2012uz}. We give detailed expressions for the usual edge-length Regge calculus as well as for its first-order versions (in face normal--connection, flux--connection and area--angle pairs of variables). Flux and area-angle variables are directly useful also in the context of LQG spin networks, spin foams and group field theory. Causal dynamical triangulations are the special case of globally constant volumes and, as such, they are also contained in this formalism. 

We conclude with an outlook on the quantization of the Laplacian operator in a quantum geometry context, and on its explicit evaluation in quantum gravity models, pointing out the difficulties that arise there.

\

Before beginning, it may be useful to tell apart original from review material.
Similar versions of exterior calculus of discrete forms are known and have been applied in 
general relativity \cite{Weingarten:1977hy, Jourjine:1987iw} 
and to other fields like random lattice field theory \cite{Itzykson:1983vc, Albeverio:1990ii}, topological field theory \cite{Adams:1996ul, Adams:1997iy, Sen:2000cr}, computational electromagnetism \cite{Gross:2004vp,Teixeira:2013ee} and computational science in general (\cite{Grady:2010wb} and references therein; reference \cite{Desbrun:2005ug} also contains a nice overview of the history of discrete calculus and further differences of the various versions in the literature). However, these versions are defined on complexes embedded in some ambient space, contrary to whay we do here.

Here, we do not rely on any embedding: while the formalism can be motivated as a discretization of functions on a triangulation of a given smooth manifold, we define it on abstract combinatorial complexes obeying the conditions of pseudo-manifolds, to comply with the use of such complexes in some quantum gravity approaches. 
A key advantage of having definitions in an abstract, combinatorial setting is a natural application of our formalism to fundamentally discrete approaches to quantum gravity.

While sections  \ref{sec1} and \ref{sec2} capitalize on the above-mentioned results (but with elements of originality we shall comment in due course), the rest of the paper contains original material. In particular, we obtain a clear picture of the role of momentum space and a systematic construction of the Laplacian with the variables of various quantum gravity models.


\section{A bra-ket formalism for discrete position spaces}\label{sec1}

In order to define a Laplacian operator, we need to have at our disposal a notion of fields, and more generally $p$-forms, in a discrete setting. Moreover, such fields have to be {\it localized} in a suitable sense, as we are dealing with a local operator and we would like to capture, through it, the local properties of the discrete geometry. In the following, we will explain the formalism in detail. For now, we just highlight the main ideas. 

At a conceptual level, as mentioned, we need some generalization of a `field at a point' in order to be able to define the action of the Laplacian on it. It is well known, in general, that for fields in the continuum the notion of position basis (exactly localized state) is unavailable. As we will see, the only existing inner product for $p$-forms involves an average (smearing) over an extended region of space. This smearing is also needed for the very definition of field theories in the continuum \cite{AlgebraicQFT}. What can be defined, in principle, is instead a basis of states restricted to a subregion of the pseudo-manifold. We will not discuss this construction in the continuum, but we will use the natural analogue of this smearing in the discrete case to define a position basis and a bra-ket formalism for discrete $p$-forms. 

From a more mathematical point of view, the definition of such basis takes the need for smearing into account together with several other structures (dualities) in both continuum and discrete geometry. In fact, the crucial point of the construction is a unification of four kinds of dualities: (1) the bra-ket duality of usual quantum mechanics in continuum position space, (2) Hodge duality on continuum Riemannian manifolds, (3) the duality of chains and cochains on complexes and finally (4) the discrete counterpart of Hodge duality constructed from the combinatorial complex and its dual. All these dualities are well known but we will take advantage of them in a novel way.


\subsection{Two dualities in the continuum}

We start with two dualitites in the continuum. The setting is thus that of a smooth manifold. Remaining in the continuum, there is no good way to unify these dualities in a strictly local manner (i.e., there is no such thing as a position basis), but a unification will be possible for their discrete analogues (the smearing being built in the discrete setting). 
\begin{enumerate}
\item[(1)] First, we have the duality of states $\phi$ in the Dirac
formalism of quantum mechanics as bras and kets \cite{Dirac:1939ck}, 
\[
\langle\phi|\quad\lolra\quad|\phi\rangle\,,
\]
where $|\phi\rangle$ is a vector in a complex Hilbert space $\mathcal{H}$
and $\langle\phi|$ is its covector, i.e., its dual linear form on
$\mathcal{H}$ with respect to the inner product $\left\langle \cdot|\cdot\right\rangle $
of $\mathcal{H}$ (which uniquely exists according to the Riesz representation
theorem). The duality is an isometric anti-isomorphism: it preserves
the norm and is linear up to complex conjugation of scalar factors.

Later, we will be interested in function spaces, and in the discrete counterpart of position space. For single particles, one has  a complete orthonormal (continuum) position basis $\{|x\rangle\}$,
\begin{eqnarray}
&\langle x|y\rangle =\delta(x,y)\,,\\
&\int\d^{d}x\,|x\rangle\langle x|  =\mathbbm{1}\,.
\end{eqnarray}
The Hilbert space $\mathcal{H}$ of such system can be identified with the square-integrable complex-valued functions $L^{2}(\R^{d},\mathbb{C})$ with inner product 
\begin{equation}
\left\langle \phi|\psi\right\rangle =\langle\phi|\int_{M}\d^{d}x\,|x\rangle\langle x|\psi\rangle=\int_{M}\d^{d}x\,\phi_{x}\psi_{x}^{*}\,,\label{eq:ScalarProductQM}
\end{equation}
where $\phi_{x}:=\left\langle \phi|x\right\rangle $ are the position basis coefficients. Thus, at the level of these position functions, the duality is just given by complex conjugation:
\[
\phi_{x}=\langle\phi|x\rangle\quad\lolra\quad\langle x|\phi\rangle=\phi_{x}^{*}\,,
\]
because of its anti-linearity.
\end{enumerate}

In the following, we are not particularly interested in quantum mechanics
but rather in a convenient notation for elements in $L^{2}$ function
spaces for ($p$-form) fields.
\begin{enumerate}
\item[(2)] Second, on a (continuum) Riemannian manifold $(M,g)$, there
is Hodge duality which maps $p$-forms $\phi\in\Omega^{p}(M)$ to
$\left(d-p\right)$-forms $\ast\phi\in\Omega^{d-p}(M)$ \cite{Nakahara:2003vq},
\begin{equation}
\fl \phi=\phi_{i_{1}\dots i_{p}}\d x^{i_{1}}\wedge\dots \wedge\d x^{i_{p}}\lolra\ast\phi=(\ast\phi)_{i_{p+1}\dots i_{d}}\d x^{i_{p+1}}\wedge\dots \wedge\d x^{i_{d}}\,,
\end{equation}
with coefficients 
\[
\fl \phi_{i_{1}\dots i_{p}}\quad\lolra\quad(\ast\phi)_{i_{p+1}\dots i_{d}}=\frac{\sqrt{g}}{\left(d-p\right)!}\epsilon_{i_{1}\dots i_{d}}g^{i_{1}j_{1}}\dots g^{i_{p}j_{p}}\phi_{i_{1}\dots i_{p}}\,.
\]
In general, it is a duality only up to a sign, $\ast\ast\phi=(-1)^{p(d-p)}\phi$. (For Lorentzian manifolds, there is an extra minus sign; this fact would be an important guiding line for extending the discrete formalism
consistently to Lorentzian geometries.) The natural inner product of $p$-forms is again an integration over position manifold by pairing a form and a dual form:
\begin{equation}
\fl \left( \phi,\psi\right) =\int_{M}\phi\wedge\ast\psi=\int_{M}\left(\phi_{i_{1}\dots i_{p}}\right)_{x}\left[(\ast\psi)_{i_{p+1}\dots i_{d}}\right]_{x}\sqrt{g_{x}}\d x^{i_{1}}\wedge\dots \wedge\d x^{i_{d}}\,.\label{eq:ScalarProductForms}
\end{equation}
This defines an $L^{2}$-space of forms $L^{2}\Omega^{p}(M)$ \cite{Rosenberg:1997to}. 
\end{enumerate}
The crucial point to notice is that this natural inner product, compatible with the tensorial ($p$-form) nature of the fields, involves an averaging (smearing) over the base manifold. Because of this tensorial structure, a simple-minded notion of a position basis is not viable, as any perfectly localized field would not be a well defined element of the above space. Only smeared fields are. We will see how the discrete setting provides a natural notion of smearing, which in turn allows us to define an analogue of a position basis even for fields.


\subsection{Exterior forms on simplicial complexes}

One can identify a natural concept of {\it discrete forms} by using a third type of duality 
\cite{Desbrun:2005ug,Albeverio:1990ii, Adams:1996ul, Teixeira:2013ee, Grady:2010wb}. For defining it, we choose finite abstract simplicial complexes as our discrete setting in contrast to the cited literature, where typically topological complexes embedded in some ambient space are the starting point.

A finite abstract {simplicial complex} $K$ (in the following, \emph{simplicial complex} for short) is a multiset of ordered subsets $\sigma$ of the set of vertices
$K_{0}=\{v_{1},v_{2},\dots ,v_{N_{0}}\}$ such that if $\sigma\in K$
and $\sigma'\In\sigma$ also $\sigma'\in K$ \cite{Kozlov:2008wc}.
In general, $\sigma'\In\sigma$ is called a face of $\sigma$. All
subsets of cardinality $p+1$ are called \emph{$p$-simplices} $\sigma_{p}\in K_{p}$
and the dimension $d$ of $K$ is defined as the maximal cardinality
of simplices in $K$. Thus, $K$ consists of $0$-simplices to $d$-simplices,
$K=\bigcup_{p=0}^d K_{p},$ and is also referred
to as a simplicial $d$-complex. The ordering of the sets $\sigma_{p}=(v_{i_{1}},\dots ,v_{i_{p}})=:(i_{1}\dots i_{p})$ defines an orientation on the complex.

\begin{enumerate}
\item[(3)] There is a duality between chains and cochains
on the simplicial complex $K$ \cite{Kozlov:2008wc}. Formal linear combinations of
$p$-simplices generate the finite vector space of \emph{$p$-chains} $c\in C_{p}(K)$
(which we take on $\mathbb{C}$) and we introduce
a bra-ket notation to write them as follows:
\[
|c\rangle=\underset{\sigma_{p}\in K_{p}}{\sum}c_{\sigma_{p}}|\sigma_{p}\rangle=\underset{\sigma_{p}\in K_{p}}{\sum}\left\langle \sigma_{p}|c\right\rangle |\sigma_{p}\rangle\,.
\]
Accordingly, linear forms on chains are called \emph{$p$-cochains} $\tilde c\in C^{p}(K)$.
As they can be expanded in the dual basis $\{\langle\sigma_{p}|\}$,
defined by the pairing $\left\langle \sigma_{p}|\sigma_{p}'\right\rangle =\delta_{\sigma\sigma'}$,
the cochain $\tilde{c}$ dual to $c$ can be written as the bra 
\[
\langle\tilde{c}|\equiv\langle c|=\underset{\sigma_{p}\in K_{p}}{\sum}c_{\sigma_{p}}^{*}\langle\sigma_{p}|=\underset{\sigma_{p}\in K_{p}}{\sum}\left\langle c|\sigma_{p}\right\rangle \langle\sigma_{p}|\,.
\]
\end{enumerate}
The connection to the first two continuum dualities is the following \cite{Desbrun:2005ug,Grady:2010wb}: on a finite triangulation of a Riemannian manifold $(M,g)$ being a geometric realization $|K|$ of an abstract simplicial complex $K$, $p$-cochains can be naturally interpreted as discretized $p$-forms $\phi\in\Omega^{p}(K)\cong C^{p}(K)$
  by smearing the continuous form $\phi_{\rm cont}\in\Omega^{p}(M)$ over $p$-surfaces $\mathcal{S}\In|K|\In M$ 
represented by chains $|\mathcal{S}\rangle=\sum_iV_{\sigma_{p}^{i}}|\sigma_{p}^{i}\rangle\in C_{p}(K)$
in the triangulation: 
\[
\phi(\mathcal{S}):= \left\langle \phi|\mathcal{S}\right\rangle = \underset{i}{\sum}V_{\sigma_{p}^{i}}\left\langle \phi|\sigma_{p}^{i}\right\rangle =\underset{i}{\sum}\int_{\sigma_{p}^{i}}\phi_{\rm cont}=\int_{\mathcal{S}}\phi_{\rm cont}\,,
\]
where $V_{\sigma_{p}}$ denotes the $p$-volume of $\sigma_{p}$ in $|K|$.
In particular, for the surface of a single $p$-simplex $\sigma_{p}$ represented by $V_{\sigma_{p}}|\sigma_{p}\rangle$, one has
\begin{equation}
\phi(\sigma_{p})=V_{\sigma_{p}}\left\langle \phi|\sigma_{p}\right\rangle =V_{\sigma_{p}}\phi_{\sigma_{p}}=\int_{\sigma_{p}}\phi_{\rm cont}.\label{eq:PrimalSmearing}
\end{equation}
Therefore, the coefficient $\phi_{\sigma_{p}}:=\langle\phi|\sigma_{p}\rangle$
has the interpretation as the averaged field value of $\phi_{\rm cont}$ over $\sigma_{p}$. Obviously, the above requires an embedding of the abstract simplicial complex into the continuum manifold in terms of a geometric realization.

However, note that, even though motivated by discretization, this definition works perfectly
well for the abstract simplicial complex $K$. We just take 
\begin{equation}
\phi_{\sigma_{p}}:=\langle\phi|\sigma_{p}\rangle\label{eq:DiscreteForm}
\end{equation}
as the definition of position coefficients of a $p$-form $\langle\phi|$. Even
a geometric interpretation in terms of $p$-volumes $V_{\sigma_{p}}$
as induced by the ambient space $M$ in the case of triangulations
is not needed at this stage, as long as we are only interested
in the forms $\langle\phi|$ themselves and not in integrated quantities
$\phi(\sigma_{p})$. 


\subsection{Choice of convention}

Before moving on to discuss the other dualities and discrete calculus, let us point out one difference between our definitions and the ones that can be found in the literature \cite{Desbrun:2005ug,Grady:2010wb}. One has in fact a choice as to where to include the geometric information encoded in
the volumes. The question is whether the $p$-volume $V_{\sigma_{p}}$
of a simplex $\sigma_{p}$ is defined explicitly in its $p$-chain
representation $V_{\sigma_{p}}|\sigma_{p}\rangle$, such that 
\[
\langle\phi|\sigma_{p}\rangle=\phi_{\sigma_{p}}=\frac{1}{V_{\sigma_{p}}}\phi(\sigma_{p})
\]
as chosen here, or whether it is already
implicit in $|\sigma_{p}\rangle$ such that 
\[\label{otco}
\langle\phi|\sigma_{p}\rangle=\phi(\sigma_{p})=V_{\sigma_{p}}\phi_{\sigma_{p}}\,,
\]
as in \cite{Desbrun:2005ug}. The former has the advantage that
the position-space measure is explicit. This is not only the usual way fields are mostly treated in physics in terms of coefficients but is especially important in the case of fields on a dynamic geometry, namely to disentangle the geometric from the field degrees of freedom.

The latter choice could be called the `math' convention since it is natural from the point of view of the mathematical properties of forms.  This is reflected in the fact that, in this convention, Hodge duality must
depend on the geometric interpretation in terms of volumes, while differentials
do not \cite{Desbrun:2005ug}. Naturally, this convention would be, in particular, useful in topological field theory (\cite{Adams:1997iy, Sen:2000cr}, where, however, a version of the topological action using the Hodge dual is eventually needed). In our choice, it is exactly the other way round (see equations (\ref{eq:HodgeDuality}) and (\ref{eq:Differential})).

There is a third convention, used in random lattice field theory \cite{Christ:1982kr,Christ:1982hv,Christ:1982bn}
by Itzykson \cite{Itzykson:1983vc}, where $\langle\phi|\sigma_{p}\rangle$
is defined as a function for every $p$, thus without volumes, but
where the duals carry the whole $d$-volume as densities. This can
be justified by the common convention in the continuum to attach the
metric part $\sqrt{g}$ of the invariant measure only to the Hodge dual forms.


\subsection{Discrete Hodge duality}\label{hod}

In order to be able to define the natural scalar product for $p$-forms also for these discrete forms, a discrete version of Hodge duality is necessary. While some approaches
\cite{Albeverio:1990ii, Adams:1996ul, Sen:2000cr, Teixeira:2013ee}
use the Whitney embedding map to define the Hodge dual which is not available for abstract complexes, in  \cite{Desbrun:2005ug} a definition is given only in terms of a dual complex (but still in a setting of embedded complex). We can take advantage of such a fourth duality also in our case of
abstract simplicial complexes under the further requirement of imposing pseudo-manifold properties.

A finite abstract \emph{simplicial pseudo $d$-manifold} is a finite abstract simplicial $d$-complex which is non-branching, strongly connected and dimensional homogeneous \cite{Gurau:2010iu}. That is, each $(d-1)$-simplex is face of exactly two $d$-simplices (non-branching), any two $d$-simplices have a strong chain of $d$-simplices
neighboring pairwise by $\left(d-1\right)$-faces (strongly connected) and every simplex is face of some $d$-simplex (dimensionally homogeneous).
\begin{enumerate}
\item[(4)] A simplicial pseudo $d$-manifold $K$ has a combinatorial
dual complex $\star K$ consisting of $(d-p)$-cells $\star\sigma_{p}$ (which we also denote as $\hat{\sigma}_{d-p}$) dual to the primal $p$-simplices $\sigma_{p}$, with orientation induced from the orientation of $K$ and cellular structure induced by the adjacency relations of $K$. The latter means that $\star\sigma\In\star\sigma'$
if, and only if, $\sigma'\In\sigma$.\footnote{The fact that every $p$-simplex contains $C_{q+1}^{p+1}={p+1\choose q+1}$ $q$-simplices translates into the condition for the dual complex to have $N_{k|l}={d+1-l \choose k-l}={d+1-l\choose d+1-k}$ $k$-cells with a given $l$-cell as a face. This property can be used as an
iterative check for constructing such dual complexes from regular graphs \cite{Bombelli:2009hg}.} Then, $\star K$ can be given as a multiset over its vertex set too.

This duality between a `primal' simplicial and a dual cell complex
induces a new type of dual chains, the chains
$\star c\in C_{p}(\star K)$ on the dual complex. This is possible because each primal
chain basis element (simplex) has a unique dual basis element. Using
Dirac notation  also for this duality, this reads 
\begin{equation}
|c\rangle=\underset{\sigma_{p}\in K}{\sum}c_{\sigma_{p}}|\sigma_{p}\rangle\quad\overset{\star}{\lolra}\quad\langle\star c|=\underset{\sigma_{p}\in K}{\sum}c_{\sigma_{p}}^{*}\langle\star\sigma_{p}|\,.\label{eq:Duality4dualchains}
\end{equation}
Note that, due to the relative orientations of the complexes, the duality holds only up to a sign (\cite{Desbrun:2005ug}, p 8; with a strong focus on the orientation properties \cite{Mattiussi:1997jp}, in \cite{Teixeira:2013ee,Teixeira:1999hv} the dual complex is even called the `twisted' complex):
\[\label{eq:sign}
\star^{2}=(-1)^{p(d-p)}.
\]
Analogously, the duality also holds between primal and dual cochains. 
\end{enumerate}
Since the Hodge dual $(d-p)$-form cannot live on $p$-simplices
but only on $(d-p)$-cells, we can regard the discrete Hodge dual of a
$p$-form $\phi\in\Omega^{p}(K)$ as its dual in the sense of this
fourth duality: 
\[
\ast\phi:=\star\phi\in\ast\Omega^{p}(K)\cong\Omega^{d-p}(\star K)\cong C^{d-p}(\star K)\,.
\]
From the above dualities, at the level of coefficients the defining condition for the
Hodge duality is the equality of the averaged field values\footnote{This is analogous to  what is done in \cite{Desbrun:2005ug}. The details of the construction differ, however, since in that work the convention \Eq{otco} is used which includes $p$-volumes.}:
\begin{equation}
(\ast\phi)_{\star\sigma_{p}}:=\phi_{\sigma_{p}}^{*}\,.\label{eq:HodgeDualDiscrete}
\end{equation}
With the bra-ket convention
\[
\left\langle \star\sigma_{p}|\phi\right\rangle =\ast\phi_{\star\sigma_{p}}\,,
\]
this can be equivalently expressed as
\[
\left\langle \star\sigma_{p}|\phi\right\rangle=\left\langle\phi|\sigma_{p}\right\rangle^{*}.
\]
In the case of triangulations discussed above, we can again analogously
view the coefficients of dual fields $\ast\phi$ as smeared fields:
\begin{equation}
\ast\phi(\star\sigma_{p})=V_{\star\sigma_{p}}(\ast\phi)_{\star\sigma_{p}}
=\int_{\hat{\sigma}_{d-p}}\ast\phi_{\rm cont}\,.
\end{equation}
Thus, one can take Hodge duality as two different perspectives to look
at the same discrete field $\phi$: either as a $p$-form $\langle\phi|$ on the primal
complex or a $(d-p)$-form $|\phi\rangle$ on the dual complex.


\subsection{Geometric interpretation of abstract complexes}

So far, we have presented a formalism for fields on abstract discrete
spaces without using any geometric information either associated directly with the simplicial complex or derived from an original continuum pseudo-manifold being discretized. However, a geometric interpretation for the elements of the simplicial complex is needed to define the inner product.

In the first place, we understand an assignment of geometry to a finite
simplicial pseudo-manifold $K$ as an assignment of $p$-volumes $V_{\sigma_{p}}$,
dual $(d-p)$-volumes $V_{\hat{\sigma}_{d-p}}$ and support $d$-volumes
$V_{\sigma_{p}}^{(d)}$ to all the simplices $\sigma_{p}$. If $K$ has a geometric realization $|K|$ in terms of a (topological) simplicial complex over a metric space, these volumes can be induced from this realization. 
In the fundamentally discrete setting of approaches to quantum gravity, on the other hand, these volumes have to be defined as functions of the geometric variables in each approach. To this end, in section 5 we will understand the simplices as locally flat and assign the volumes according to the functions of geometric variables one obtains in the case of a geometric realization on flat space. Therefore, we now discuss this case in detail.

\begin{figure}
\begin{centering}
a)
\includegraphics[scale=0.3]{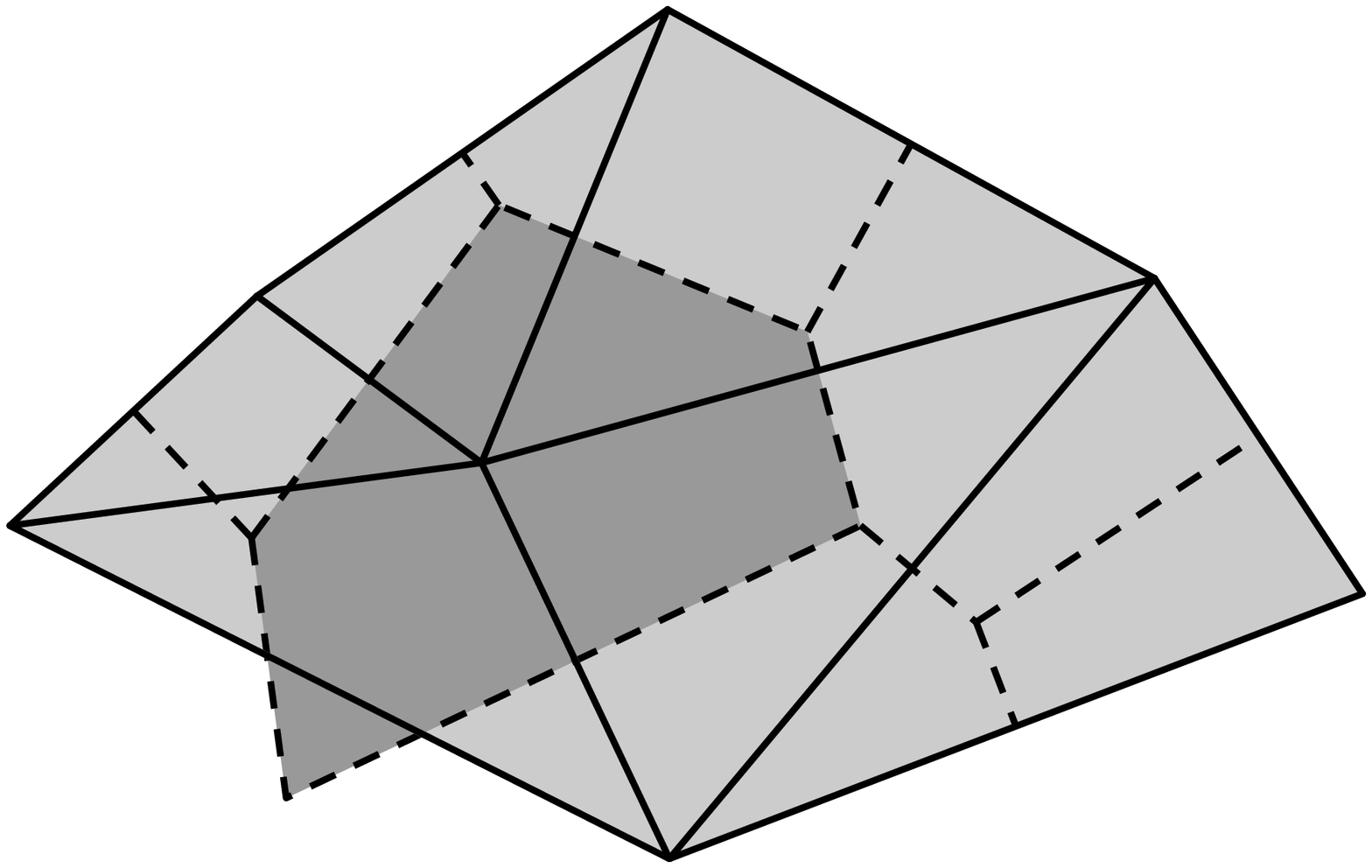}
b)
\includegraphics[scale=0.3]{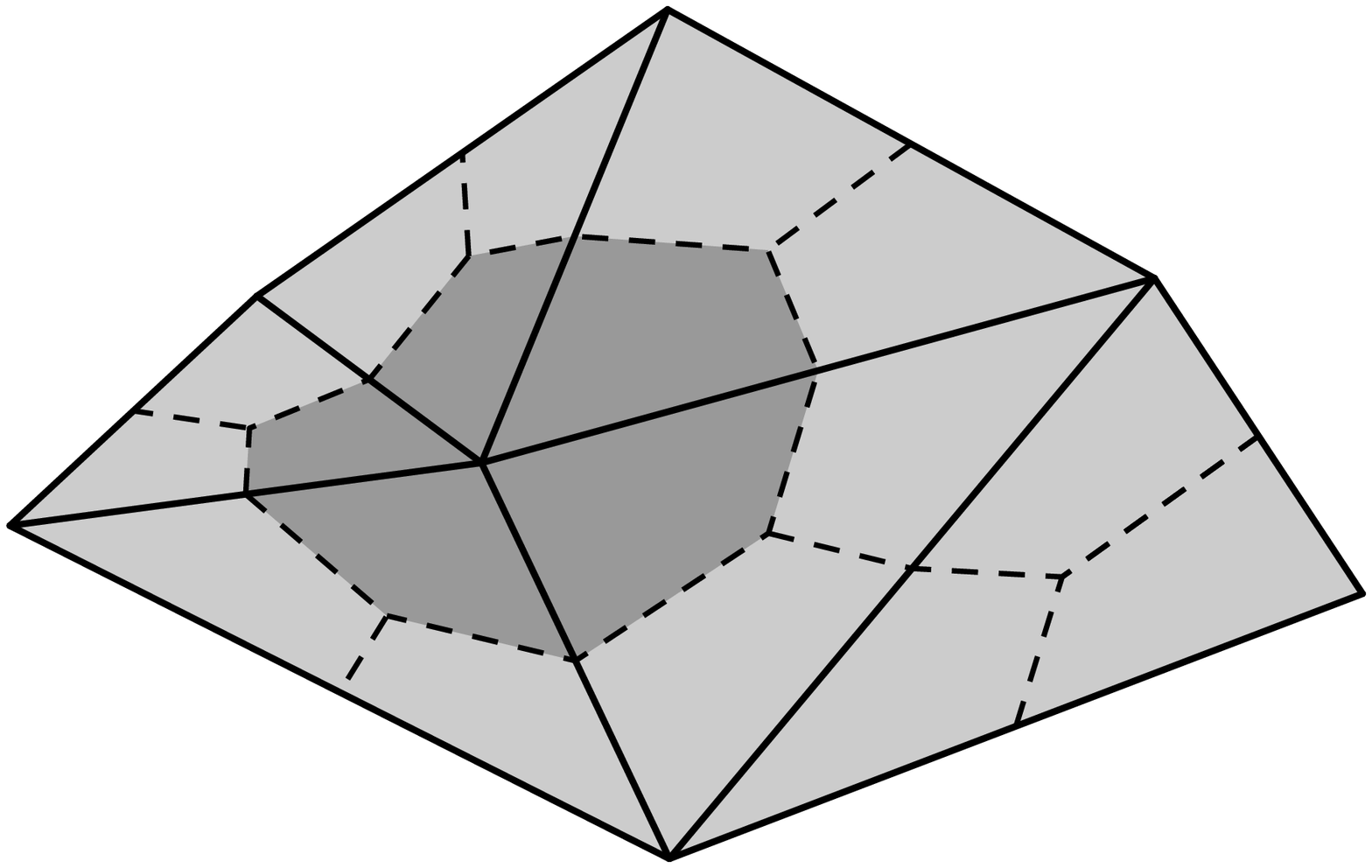} 
\par\end{centering}
\caption{\label{fig1} Circumcentric $(a)$ and barycentric $(b)$ dual cells to the same simplicial $d=2$ complex; for the purpose of illustration, dual edges are dashed and one dual 2-cell is highlighted.}
\end{figure}

While the primary volumes can be taken directly from a geometric realization, dual and support volumes depend on how the dual complex is realized, i.e., how it is concretely constructed from (or embedded into) the primal complex. The most common choices in the literature are circumcentric  \cite{Desbrun:2005ug} and barycentric \cite{Albeverio:1990ii, Adams:1996ul, Teixeira:2013ee} dual complexes (figure \ref{fig1}).

For constructing the \emph{circumcentric} dual, one chooses the circumcenters
of the $d$-simplices as the $0$-cells and builds up higher cells (i.e., with $p\geq 1$)
connecting them according to the combinatorics induced from the primal
complex; sub-cells of the dual complex are then automatically identified as well.

In the case of a geometric realization in terms of a Delaunay triangulation, the circumcentric dual complex is a Voronoi decomposition. A \emph{Delaunay triangulation} is obtained by constructing $d$-simplices  from a set of points in a metric space such that no point is in the interior of the circumsphere of any $d$-simplex. From the same set of points, a $d$-cell of a \emph{Voronoi decomposition} associated with some point $P$ is constructed as the set of points closer to $P$ than to any other in the set.

\begin{figure}
\begin{centering}
$(a)$
\includegraphics[scale=0.3]{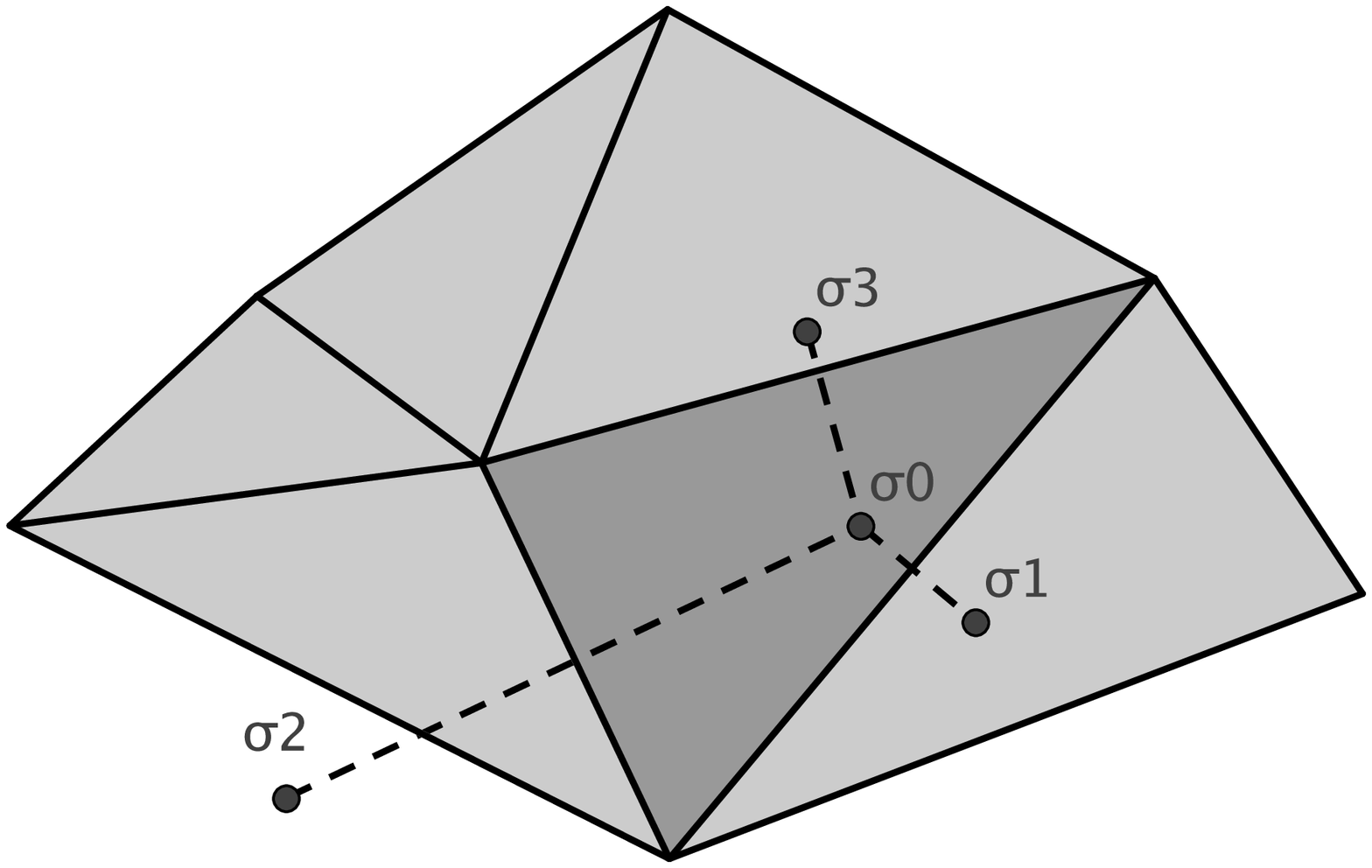}
$(b)$
\includegraphics[scale=0.3]{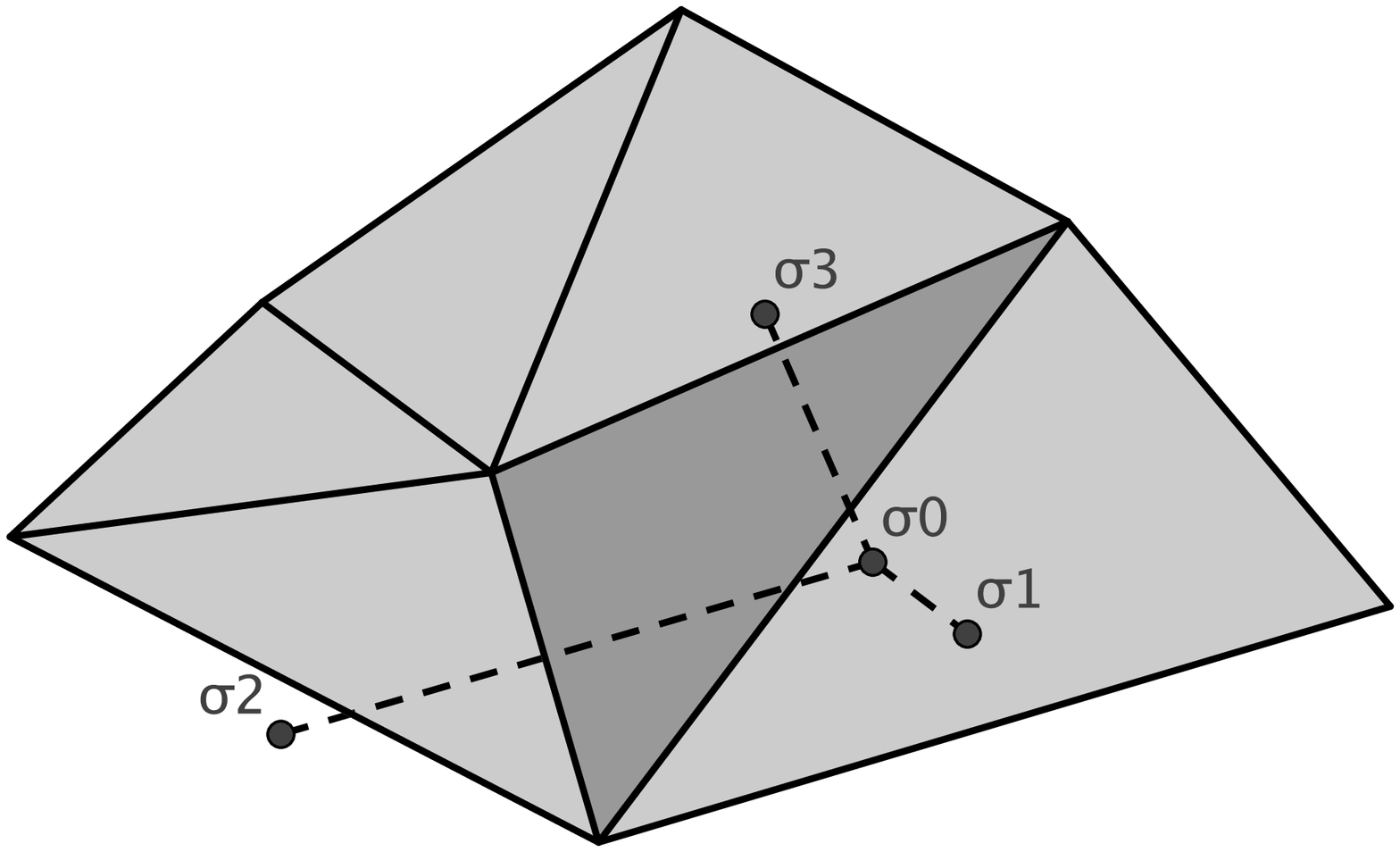}

$(c)$
\includegraphics[scale=0.3]{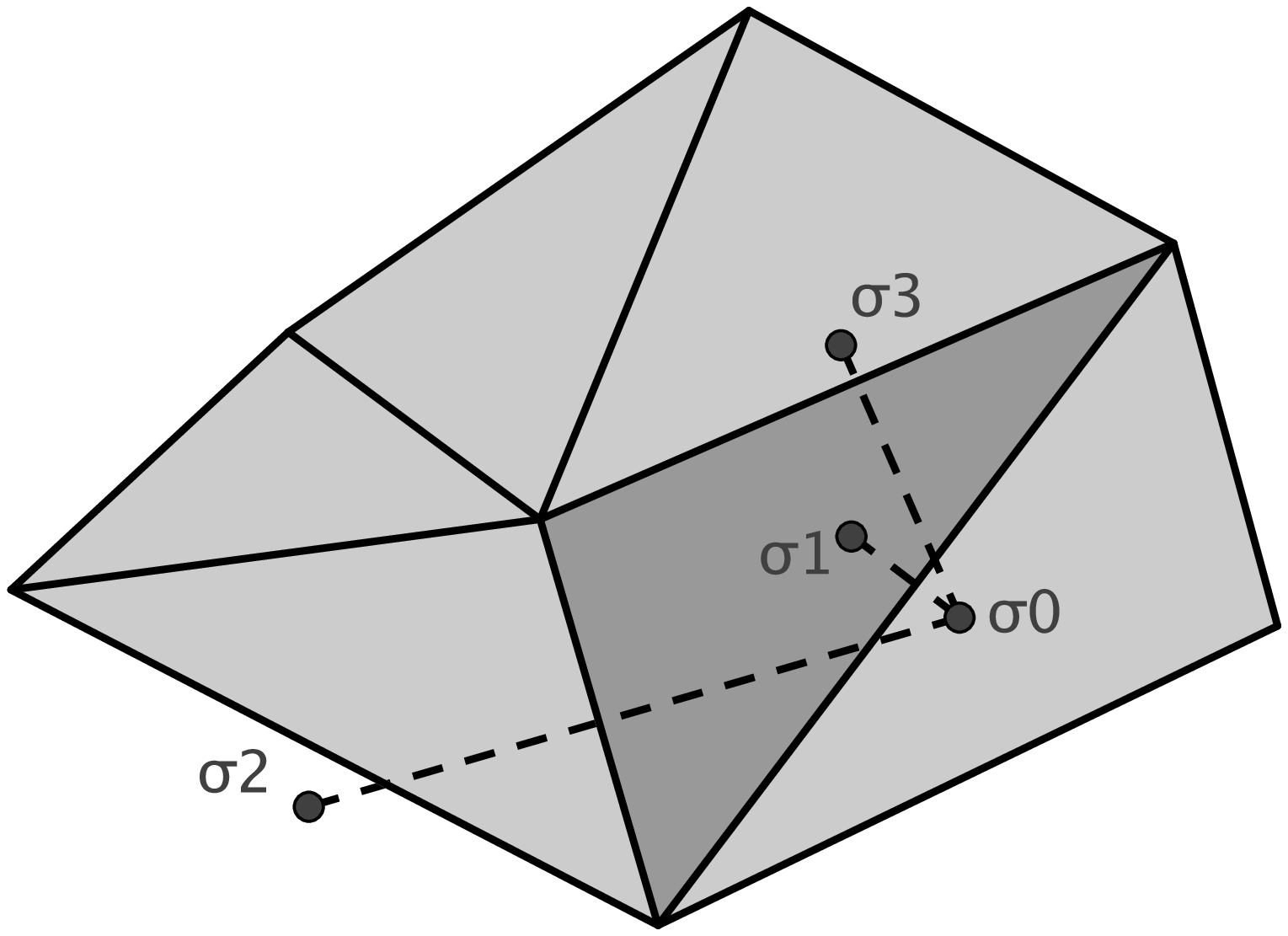}
$(d)$
\includegraphics[scale=0.3]{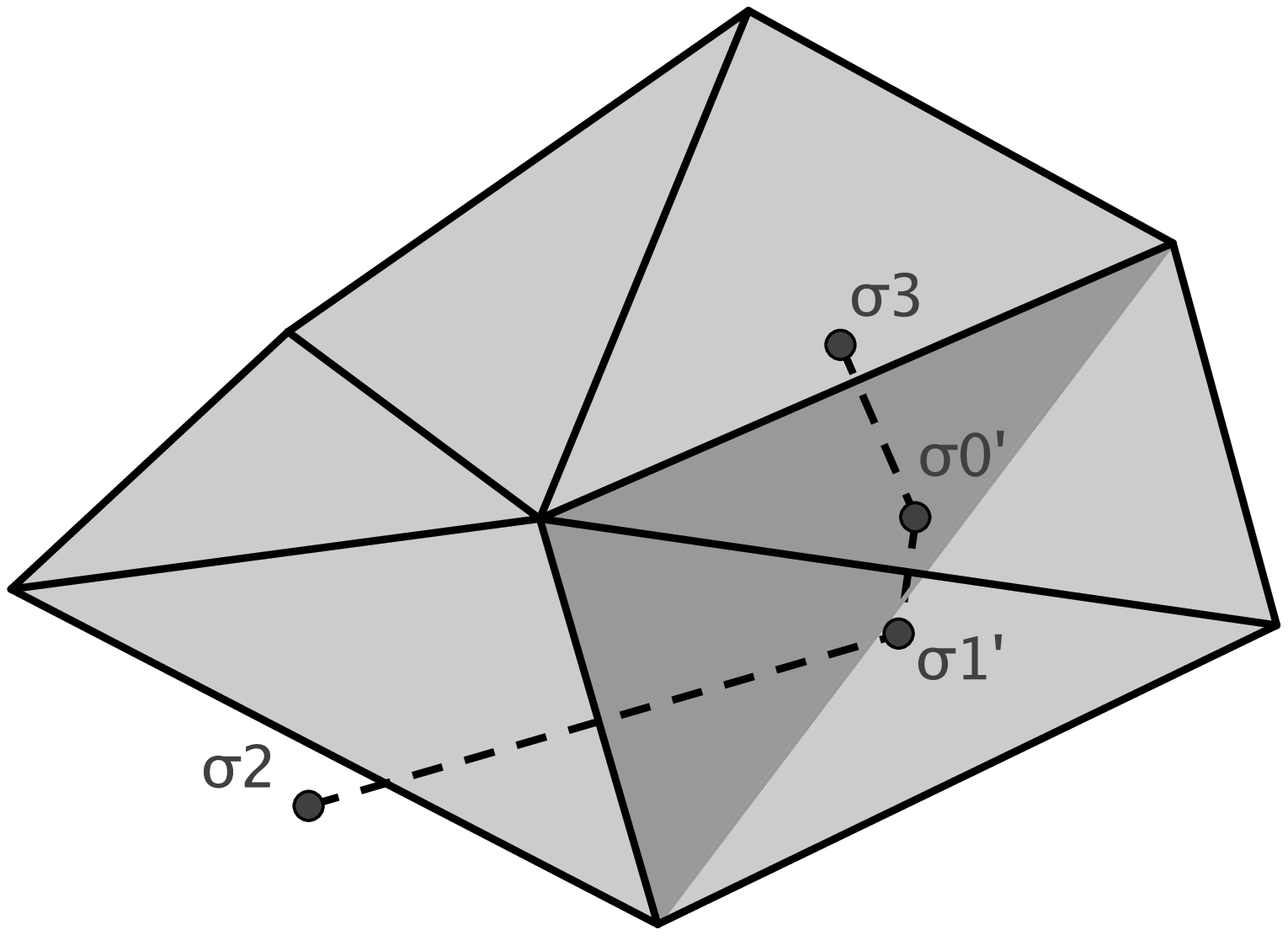} 
\par\end{centering}

\caption{\label{fig2}Examples of the dual edges to the the faces of a triangle in a simplicial $d=2$ complex: In the first picture $(a)$, the dual edges are the sum of the distance of the dual vertices $\hat{\sigma}_i$ to the face. In the second $(b)$, the dual vertex $\hat{\sigma}_0$ lies outside the triangle. Its distance to the face line therefore has to be subtracted, but the dual length $\hat{l}_{\hat{\sigma}_0 \hat{\sigma}_1}$ is still positive. In the third case $(c)$, both the vertices of  $\hat{l}_{\hat{\sigma}_0 \hat{\sigma}_1}$ lie outside their simplices such that  $\hat{l}_{\hat{\sigma}_0 \hat{\sigma}_1}$ is negative. Exactly when this happens, the triangulation cannot be Delaunay because the circumcenter is closer to the neighbor than to its own triangle. The forth picture $(d)$ shows the Delaunay triangulation for the primal points of $(c)$, where the triangle originally considered, and thus its dual vertex, does not exist (though still shaded for comparison).}
\end{figure}

For this reason, the circumcentric dual is often also called Voronoi dual. But this is meaningful only for Delaunay triangulations. For an arbitrary triangulation, the circumcentric dual complex and the Voronoi decomposition with respect to the vertex set of the triangulation are different. In fact, the Voronoi decomposition does not have the structure of a dual complex for triangulations which are not Delaunay. This is particularly important in the abstract setting where the simplicial pseudo-manifold is considered as a gluing of $d$-simplices and the geometry of each is to be defined independently of its neighbors.   The difference is further detailed in the discussion of the Laplacian in section \ref{lapr} and in figure \ref{fig2}.

The \emph{barycentric} dual, on the other hand, is defined by a barycentric
subdivision of all simplices, assumed to be flat in their interior. The barycenters of $d$-simplices define
the dual points, and metrically connecting them iteratively to the
barycenters of lower simplices defines the realization of the higher
cells.

While the circumcentric dual is not {built} by a (circumcentric)
subdivision, a simplicial subdivision nevertheless can be constructed analogously to the barycentric subdivision \cite{Desbrun:2005ug}. One can
therefore in general define the support volumes $V_{\sigma_{p}}^{(d)}$
to be the volume of the symmetric difference of all $d$-simplices in this subdivision
having $\sigma_{p}$ (or equivalently $\star\sigma_{p}$) on its boundary. (To account also for the case of circumcentric dual complexes with some circumcenters outside their simplices, the symmetric difference instead of the union \cite{Desbrun:2005ug} of simplices in the subdivision has to be used. This can also happen in the case of primal Delaunay triangulations.) Since the whole of support volumes for
a given $p$ takes all the simplices in the subdivision into account,
they indeed define a space measure summing up to the total volume
$V$ of the complex:
\[
\underset{\sigma_{p}\in K_{p}}{\sum}V_{\sigma_{p}}^{(d)}=V\,.
\]
In the circumcentric case, the support volumes are not independent
of the $p$-volumes but proportional to their product \cite{Desbrun:2005ug}:
\begin{equation}
V_{\sigma_{p}}^{(d)}=\frac{1}{d}V_{\sigma_{p}}V_{\hat{\sigma}_{d-p}}\,.\label{eq:VolProduct}
\end{equation}
We will give explicit expressions of these volumes in terms of various
geometric variables below. 


\subsection{Inner product and position-space measure}

With a well-defined meaning given to bras and kets of discrete fields,
only a slight modification to the forth duality (\ref{eq:Duality4dualchains})
is needed in order to have the geometric $L^{2}$ inner product
on the simplicial complex $K$ analogous to the continuum case (again analogous to \cite{Desbrun:2005ug} but different in details of convention): 
\[
\langle\phi|\psi\rangle:=\underset{\sigma_{p}}{\sum}V_{\sigma_{p}}^{(d)}\phi_{\sigma_{p}}\psi_{\star\sigma_{p}}^{*}=\langle\phi|\underset{\sigma_{p}}{\sum}V_{\sigma_{p}}^{(d)}|\sigma_{p}\rangle\langle\star\sigma_{p}|\psi\rangle\,,
\]
where we took a position-space measure into account in terms of the $d$-volumes
$V_{\sigma_{p}}^{(d)}$ associated with the pairs of primal and dual
$p$-simplices. This inner product is obtained by a resolution of the identity
\[
\underset{\sigma_{p}}{\sum}V_{\sigma_{p}}^{(d)}|\sigma_{p}\rangle\langle\star\sigma_{p}|=\mathbbm{1}\,,
\]
which 
in our convention demands, for reasons of consistency, a modification of the pairing
of primal and dual chains:
\[
\langle\star\sigma_{p}|\sigma_{p}'\rangle:=\frac{1}{V_{\sigma_{p}}^{(d)}}\delta_{\sigma\sigma'}\,.
\]
By the third duality between chains and cochains, this directly yields
the same form of completeness and orthonormality relations for primal
and dual cochains. While the $p$-volumes in (\ref{eq:PrimalSmearing}) and (\ref{eq:HodgeDualDiscrete})
are not needed to define the field space, the position measure $V_{\sigma_{p}}^{(d)}$
is crucial and it is at this stage where a geometric interpretation
is needed.

For the inner product to be well defined, the space of $p$-form fields does not have to be constrained further. Since its dimension is the number of $p$-simplices in the finite complex $K$, 
\[\label{dimo}
\dim\Omega^{p}(K)=\dim\Omega^{d-p}(\star K)={\rm Card}(K_{p})<\infty\,,
\]
the field space $\Omega^{p}(K)\cong\Omega^{d-p}(\star K)$ is already
the discrete $L^{2}$ space.


\subsection{The bra-ket formalism}

To define a formalism with unique types of bras and kets, we now go
one step further (beyond \cite{Desbrun:2005ug}) and identify primal
chains with dual cochains and dual chains with primal cochains:
\begin{equation}
|\sigma_{p}\rangle\equiv|\star\sigma_{p}\rangle\,,\qquad \langle\sigma_{p}|\equiv\langle\star\sigma_{p}|\,.
\end{equation}
We end up having just one complete orthonormal basis (normed to the inverse volume factors):
\begin{eqnarray}
\left\langle \sigma_{p}|\sigma'_{p}\right\rangle =\frac{1}{V_{\sigma_{p}}^{(d)}}\delta_{\sigma\sigma'}\,,\\
\underset{\sigma_{p}}{\sum}V_{\sigma_{p}}^{(d)}|\sigma_{p}\rangle\langle\sigma_{p}|=\mathbbm{1}\,.
\end{eqnarray}
With this identification, we can now write the Hodge dual (\ref{eq:HodgeDualDiscrete})
as
\begin{equation}
\left\langle \ast\phi|\sigma_{p}\right\rangle :=\left\langle \star\sigma_{p}|\phi\right\rangle \equiv \left\langle \sigma_{p}|\phi\right\rangle =\left\langle \phi|\sigma_{p}\right\rangle ^{*}.\label{eq:HodgeDuality}
\end{equation}
In the case of ket fields, one has to be careful in defining such a notation because of the sign (\ref{eq:sign}):
\begin{equation}
\left\langle \sigma_{p}|\ast\phi\right\rangle :=\left\langle \phi|\star\star\sigma_{p}\right\rangle = (-1)^{p(d-p)}\left\langle \phi|\sigma_{p}\right\rangle =(-1)^{p(d-p)}\left\langle \sigma_{p}|\phi\right\rangle ^{*}.\label{eq:HodgeDuality2}
\end{equation}
In this way, the sign factor in the duality of complexes induces consistently the usual sign factor in the Hodge duality:
\[
\left\langle \ast\ast\phi|\sigma_{p}\right\rangle =
\left\langle \phi|\star\star\sigma_{p}\right\rangle =
(-1)^{p(d-p)}\left\langle \phi|\sigma_{p}\right\rangle .
\]
The following commutative diagram shows the identifications and dualities
by which the discrete $L^{2}$ position function space is defined:
\begin{eqnarray}
\xymatrix{\Omega^{p}(K)\ar@{<->}[d]^{\cong}\ar@{<->}[r]^{\ast} & \Omega^{d-p}(\star K)\ar@{<->}[d]^{\cong}\\
C^{p}(K)\ar@{<->}[rd]^{\sim}\ar@{<->}[r]^{\star}\ar@{<->}[d]^{\equiv} & C^{d-p}(\star K)\ar@{<->}[d]^{\equiv}\\
C_{d-p}(\star K)\ar@{<->}[ru]_{\ \ \ \sim}\ar@{<->}[r]_{\star} & C_{p}(K)
}\nonumber\\
\end{eqnarray}
All the maps are well known \cite{Desbrun:2005ug,Grady:2010wb} except
for the last identification denoted as `$\equiv$', which makes it possible to have a
Dirac position-space notation for arbitrary $p$-fields on simplicial
pseudo-manifolds with an assigned set of  geometric data.

Note that, since by the last identification the pairing of the chain-cochain
duality is modified, too, the fields finally have bra and ket component
expansion 
\[
\langle\phi|=\underset{\sigma_{p}\in K}{\sum}V_{\sigma_{p}}^{(d)}\phi_{\sigma_{p}}\langle\sigma_{p}|\quad\overset{*}{\longleftrightarrow}\quad|\phi\rangle=\underset{\sigma_{p}\in K}{\sum}V_{\sigma_{p}}^{(d)}\phi_{\sigma_{p}}^{*}|\sigma_{p}\rangle\,.
\]
As an example, a field living on the $d$-simplices represented by chains $|\sigma_d \rangle\in C_d(K)$, i.e. a primal $d$-form field $\phi\in\Omega^d(K)$, has an expansion in terms of the cochain basis elements $\langle\sigma_d|\in C^d(K)$ with an explicit volume measure $V_{\sigma_d}^{(d)}=V_{\sigma_d}$, $\langle\phi|=\sum V_{\sigma_d} \phi_{\sigma_d}\langle\sigma_d|$. Its Hodge dual is a scalar on the dual vertices represented by chains $\langle\star\sigma_d|=\langle\hat\sigma_0|\in C_0(\ast K)$, identified with primal cochains $\langle\hat\sigma_0| \equiv \langle\star\sigma_d|$, that is a dual 0-form $\star\phi\in\Omega^0(\star K)$, having an expansion with a trivial vertex volume measure in dual cochains $\langle\hat\sigma_0|\in C^0(\star K)$ which can be identified with primal $d$-chains, $|\phi\rangle=\sum\phi_{\sigma_d}^*|\sigma_d\rangle$.


\section{Laplacian on simplicial pseudo-manifolds}\label{sec2}

In order to define the Laplacian, we have first to introduce
discrete calculus on complexes by defining a differential. Then, the
formal expression of the Hogde Laplacian is well defined on simplicial
pseudo-manifolds and we can analyze its properties in the case of dual scalar
fields.


\subsection{Exterior calculus on complexes}

The exterior differential operator on discrete forms is constructed by taking the Stokes theorem as a definition \cite{Desbrun:2005ug,Grady:2010wb}.
For the integration of the differential of a form $\phi\in\Omega^{p-1}(K)$
over one simplex $\sigma_{p}$ in the triangulation of a pseudo-manifold
with corresponding complex $K$, the theorem states that 
\[
\d\phi(\sigma_{p})=\int_{\sigma_{p}}\d\phi_{\rm cont}=\int_{\partial\sigma_{p}}\phi_{\rm cont}=\phi(\partial\sigma_{p})\,.
\]
Therefore, we define the \emph{differential} of $\phi\in\Omega^{p-1}(K)$
on an abstract simplicial complex $K$ as\footnote{The differential operator is just a modified version of the coboundary operator, which is the operator adjoint to the boundary operator with respect to the third duality between chains and cochains. It is modified because, in the convention chosen here, we have to explicitly keep track of the volume factors. In the math convention \cite{Desbrun:2005ug}, on the other hand, the differential is exactly the coboundary operator.}
\begin{equation}
\fl \d\phi(\sigma_{p})=V_{\sigma_{p}}\left\langle \d\phi|\sigma_{p}\right\rangle :=\phi(\partial\sigma_{p}):=\underset{\sigma_{p-1}\in\partial\sigma_{p}}{\sum}\mbox{sgn}(\sigma_{p-1},\sigma_{p})V_{\sigma_{p}}\left\langle \phi|\sigma_{p-1}\right\rangle\,.\label{eq:Differential}
\end{equation}
The sign factor takes into account the orientation of the faces $\sigma_{p-1}=(i_{1}\dots\hat{i}_{j}\dots i_{p})$
relatively to the bulk simplex $\sigma_{p}=(i_{1}\dots i_{p})$ via the
permutation of their vertices: 
\[
\mbox{sgn}(\sigma_{p-1},\sigma_{p}):=\sgn(i_{1}\dots\hat{i}_{j}\dots i_{p})\,\sgn(i_{1}\dots i_{p})\,.
\]
Similarly, the differential on dual forms $\phi\in\Omega^{d-p-1}(\star K)\cong\Omega^{p+1}(K)$ can be defined as 
\[
\fl V_{\hat{\sigma}_{d-p}}\left\langle \hat{\sigma}_{d-p}|\d\phi\right\rangle :=\underset{\hat{\sigma}_{d-p-1}\in\partial\hat{\sigma}_{d-p}}{\sum}\sgn(\hat{\sigma}_{d-(p+1)},\hat{\sigma}_{d-p})\,V_{\hat{\sigma}_{d-(p+1)}}\left\langle \hat{\sigma}_{d-(p+1)}|\phi\right\rangle
\]
or equivalently 
\[
V_{\star\sigma_{p}}\left\langle \sigma_{p}|\d\phi\right\rangle :=\underset{\sigma_{p+1};\sigma_{p}\in\partial\sigma_{p+1}}{\sum}\mbox{sgn}(\sigma_{p+1},\sigma_{p})\,V_{\star\sigma_{p+1}}\left\langle \sigma_{p+1}|\phi\right\rangle\,.
\]
One can easily check that indeed the differential on the dual complex is the adjoint
to the differential on the primal one, $\left\langle \d\phi|\psi\right\rangle =\left\langle \phi|\d\psi\right\rangle$. More precisely, if we do not write the inner product directly as
a pairing of bra and ket but as a bilinear form on either $\Omega^{p}(K)$
or $\Omega^{d-p}(\star K)$, the adjoint operator of the differential
as usual is
\[
\delta:=(-1)^{d(p+1)+1}\ast\d\ast\,,
\]
taking into account the sign of multiple Hodge operations \cite{Desbrun:2005ug}.


\subsection{Laplacian on dual scalar fields and its properties}\label{lapr}

Using the above notions of discrete differential and codifferential, we can now simply define the discrete Laplacian using the standard definition of the Hodge--Laplace--Beltrami operator in the well-known form \cite{Rosenberg:1997to}
\[\label{Delta}
\boxd{\Delta_{p}:=\Delta=\delta\d+\d\delta\,,}
\]
which has now a well-defined meaning on arbitrary $p$-forms on a simplicial
pseudo-manifold. In particular, we are interested in the action of this Laplacian
on dual scalar fields $\phi\in\Omega^{0}(\star K)\cong\Omega^{d}(K)$,
that is, fields living on $d$-simplices\footnote{In the literature of Regge calculus, a Laplacian of the same form is derived for a primal scalar field (i.e., a scalar field living on the vertices of the primal simplicial complex) in the circumcentric case \cite{Hamber:2009zz}. Then, the dual Laplacian $\Delta_{d}$ is guessed to have exactly the form (\ref{eq:ScalarLaplace}).}:
\begin{eqnarray}
\left(-\Delta_{d}\phi\right)_{\hat{\sigma}_{0}} & =-\langle \hat{\sigma}_{0}|(-1)^{d(1+1)+1}\ast\d\ast\d\phi\rangle \nonumber \\
 & =(-1)^{d(d-d)}\left\langle \d\ast\d\phi|\sigma_{d}\right\rangle \nonumber \\
 & =\frac{1}{V_{\sigma_{d}}}\underset{\sigma_{d-1}\in\partial\sigma_{d}}{\sum}\sgn(\sigma_{d-1},\sigma_{d})V_{\sigma_{d-1}}\left\langle \ast\d\phi|\sigma_{d-1}\right\rangle \nonumber \\
 & =\frac{1}{V_{\sigma_{d}}}\underset{\sigma_{d-1}\in\partial\sigma_{d}}{\sum}\sgn(\sigma_{d-1},\sigma_{d})V_{\sigma_{d-1}}\left\langle \hat{\sigma}_{1}|\d\phi\right\rangle \nonumber \\
 & =\frac{1}{V_{\sigma_{d}}}\underset{\sigma_{d-1}\in\partial\sigma_{d}}{\sum}\sgn(\sigma_{d-1},\sigma_{d})\frac{V_{\sigma_{d-1}}}{V_{\hat{\sigma}_{1}}}\underset{\hat{\sigma}_{0}\in\partial\hat{\sigma}_{1}}{\sum}\sgn(\hat{\sigma}_{0},\hat{\sigma}_{1})\left\langle \hat{\sigma}_{0}|\phi\right\rangle \nonumber \\
 & =\frac{1}{V_{\sigma_{d}}}\underset{\sigma'_{d}\sim\sigma_{d}}{\sum}\frac{V_{\sigma_{d}\cap\sigma_{d}'}}{V_{\star\left(\sigma_{d}\cap\sigma_{d}'\right)}} \left(\phi_{\hat{\sigma}_{0}}-\phi_{\hat{\sigma}'_{0}}\right)\label{eq:ScalarLaplace}\,.
\end{eqnarray}
In the first line, the usual vanishing of $\delta\propto\ast\d\ast$ on 0-forms is used, while in the next four lines the differential and Hodge star operator are applied one after the other. The last line is just a reordering of terms.
The dual volumes $V_{\star\left(\sigma_{d}\cap\sigma_{d}'\right)}$ in the denominator are the lengths of the dual edges between dual points $\hat{\sigma}_{0}$ and $\hat{\sigma}'_{0}$, and we write them as $\hat{l}_{\sigma\sigma'}=V_{\star\left(\sigma\cap\sigma'\right)}$ (suppressing from now on the dimension index in $\sigma=\sigma_{d}$). Thus, the action of the Laplacian on a scalar field ket is of the general type of a graph Laplace matrix \cite{Chung:1997tk}:
\begin{eqnarray}
-\Delta_{d}|\phi\rangle & =\underset{\sigma}{\sum}V_{\sigma}|\sigma\rangle\left\langle \sigma|\Delta_{d}\phi\right\rangle\nonumber \\
 & =\underset{\sigma}{\sum}|\sigma\rangle\underset{\sigma'\sim\sigma}{\sum}\frac{V_{\sigma\cap\sigma'}}{\hat{l}_{\sigma\sigma'}}\big(\left\langle \sigma|\phi\right\rangle -\left\langle \sigma'|\phi\right\rangle \big)\nonumber\\
 & =\left[\underset{\sigma}{\sum}\left(\underset{\sigma'\sim\sigma}{\sum}w_{\sigma\sigma'}\right)|\sigma\rangle\langle\sigma|\right]|\phi\rangle-\left(\underset{\sigma}{\sum}\underset{\sigma'\sim\sigma}{\sum}w_{\sigma\sigma'}|\sigma\rangle\langle\sigma'|\right)|\phi\rangle\nonumber\\
 & =: D|\phi\rangle-A|\phi\rangle\,.\label{DAeq}
\end{eqnarray}
On the 1-skeleton graph of the dual complex, it is a difference of an off-diagonal adjacency matrix $A$ in terms of weights
\[
w_{\sigma\sigma'}:=\frac{V_{\sigma\cap\sigma'}}{\hat{l}_{\sigma\sigma'}}\label{weights}
\]
and a diagonal degree matrix $D$ with entries 
\[
D_{\sigma\sigma}=\sum_{\sigma'\sim\sigma}w_{\sigma\sigma'}\label{degree}.
\]
The Laplace matrix position elements $\left(-\Delta_{d}\phi\right)_{\sigma}$, on the other hand, come with an additional inverse volume and are
\[
\frac{w_{\sigma\sigma'}}{V_{\sigma}}\label{lapelements}.
\]
By definition, such discrete (graph) Laplacians obey three desirable properties \cite{Chung:1997tk,Wardetzky:2008kk}: 
\begin{enumerate}
\item[1.] \emph{Null condition}. $\left(\Delta_d\phi\right)=0$ if, and only if, $\phi$
is constant. This is obvious because $\Delta_d\phi$ is the difference of position values of $\phi$. The zero mode of the spectrum of $\Delta_d$
reflects the fact that $K$ corresponds to a closed pseudo-manifold. 
\item[2.] \emph{Self-adjointness}. The Laplace operator is self-adjoint
with respect to the inner product 
\[
\left\langle \phi|\Delta_d\psi\right\rangle =\left\langle \Delta_d\phi|\psi\right\rangle\,.
\]
This is reflected by the symmetry of its weights $w_{\sigma\sigma'}$,
though at the level of position coefficients $\left(\Delta_d\phi\right)_{\sigma}$, equation (\ref{eq:ScalarLaplace}), the inverse-volume factor $V_{\sigma}^{-1}$ spoils this symmetry. 
\item[3.] \emph{Locality}. The action of $\Delta_d$ at any given position, $\left(\Delta_d\phi\right)_{\sigma}$,
is not affected by field values $\phi_{\sigma'}$ at non-neighboring
positions $\sigma'\nsim\sigma$. In discrete calculus, this comes directly
from the definition of the Laplacian as a second-order differential
operator. 
\end{enumerate}
In the case of a simplicial decomposition $|K|$ of a pseudo-manifold $M$,
a further natural condition which is built into the formalism from
the start (by the definition of differentials via the Stokes theorem) is
the following:
\begin{enumerate}
\item[4.] Convergence to the continuum Laplacian under refinement of triangulations. 
\end{enumerate}
To see this, consider a region $\Omega\in M$ large compared
to the scale $a\sim\left(V_{\sigma_{p}}\right)^{\frac{1}{p}}$ of
simplices $\sigma_{p}\in K$, in which the function $\phi$ and its
derivatives do not vary strongly. Using equation (\ref{eq:VolProduct}),
$V_{\sigma\cap\sigma'} V_{\star\left(\sigma\cap\sigma'\right)}\approx V_{\sigma\cap\sigma'}^{(d)}d$, we have 
\begin{eqnarray}
\underset{\sigma\in\Omega}{\sum}V_{\sigma}(-\Delta_{d}\phi)_{\sigma} & =\underset{\sigma\in\Omega}{\sum}\,\underset{\sigma'\sim\sigma}{\sum}\frac{V_{\sigma\cap\sigma'}}{\hat{l}_{\sigma\sigma'}}\left(\phi_{\sigma}-\phi_{\sigma'}\right)\approx d\underset{\sigma\in\Omega}{\sum}\,\underset{\sigma'\sim\sigma}{\sum}V_{\sigma\cap\sigma'}^{(d)}\frac{\phi_{\sigma}-\phi_{\sigma'}}{\hat{l}_{\sigma\sigma'}^{2}}\nonumber\\
 & \approx2d\ \mbox{Vol}(\Omega)\underset{\hat{\sigma}_{1}\in\Omega}{\sum}\frac{\phi_{\sigma}-\phi_{\sigma'}}{a^{2}}\,.\label{ContLimit1}
\end{eqnarray}
Summing over all the dual edges $\hat{\sigma}_{1}\in\Omega$ gives
effectively a rotationally invariant expression. In particular, it
is an average over hypercubic lattices and the difference term can
readily be seen to be the Laplacian in the continuum limit, just as in standard lattice
field theory with hypercubic lattice size $a$. As $\phi_{\sigma+ae_{\mu}}\underset{a\ra0}{\longrightarrow}\phi_{\sigma}+a\left(\partial^{\mu}\phi\right)_{\sigma}e_{\mu}+\mathcal{O}(a^{2})$, the difference term gives 
\begin{eqnarray}
\fl \overset{2d}{\underset{\sigma'}{\sum}}\frac{\phi_{\sigma}-\phi_{\sigma'}}{a^{2}} & =-\overset{d}{\underset{\mu=1}{\sum}}\frac{1}{a}\left(\frac{\phi_{\sigma+ae_{\mu}}-\phi_{\sigma}}{a}-\frac{\phi_{\sigma}-\phi_{\sigma-ae_{\mu}}}{a}\right)\nonumber\\
 & \underset{a\ra0}{\longrightarrow}-\overset{d}{\underset{\mu=1}{\sum}}\frac{\left(\partial_{\mu}\phi\right)_{\sigma}-\left(\partial_{\mu}\phi\right)_{\sigma-ae_{\mu}}}{a}e^{\mu}\approx-\overset{d}{\underset{\mu=1}{\sum}}\left(\partial^{\mu}\partial_{\mu}\phi\right)_{\sigma}\,.\label{ContLimit2}
\end{eqnarray}
Despite the validity of the above properties, one has to expect that it is not possible to preserve all the features of the continuum Laplacian in the discrete setting. This has
been shown in the case of two-dimensional triangulations \cite{Wardetzky:2008kk}. As a result, the definition of a discrete counterpart of the continuum Laplacian cannot be unique. In our case, it is therefore natural to wonder which properties of the continuum Laplacian are not preserved by the discrete Laplacian $\Delta_{d}$.

The answer turns out to depend also on the specific choice of the geometry of the dual complex, that is, on the choice of its geometric embedding into the primal complex. The two distinguishing features are linear precision and positivity. 
\begin{enumerate}
\item[5.] \emph{Linear precision}. $(\Delta_{d}\phi)_{\sigma}=0$ for straight-line
triangulations $|K|$ of flat space $M\In \mathbb{R}^{d}$ and linear functions
$\phi(x^{\mu})=c+\sum_{i=1}^d c_{\mu}x^{\mu}$
in Cartesian coordinates $x^{\mu}$. By linearity, this is equivalent
to a vanishing Laplacian $(\Delta_{d}x)_{\sigma}=0$ of the coordinate
field $x$ (considered as a bunch of scalars $x^{\mu}$). 
\end{enumerate}
Linear precision holds for circumcentric dual geometries, in which case the dual
lengths are $\hat{l}_{\sigma\sigma'}=\left|x_{\hat{\sigma}}-x_{\hat{\sigma}'}\right|$
and (with unit face normals $\hat{n}_{\sigma\sigma'}=$ $\frac{x_{\hat{\sigma}}-x_{\hat{\sigma}'}}{\left|x_{\hat{\sigma}}-x_{\hat{\sigma}'}\right|}$) 
\[
(\Delta_{d}x)_{\sigma}\sim\underset{\sigma'\sim\sigma}{\sum}\frac{V_{\sigma\cap\sigma'}}{\hat{l}_{\sigma\sigma'}}\left(x_{\hat{\sigma}}-x_{\hat{\sigma}'}\right)=\underset{\sigma'\sim\sigma}{\sum}V_{\sigma\cap\sigma'}\hat{n}_{\sigma\sigma'}=0
\]
is true because these are exactly the closure conditions for the polyhedron
$\sigma$. This property fails, on the other hand, for the barycentric case. One could heuristically understand this by noting that generically $\hat{l}_{\sigma\sigma'}\ne\left|x_{\hat{\sigma}}-x_{\hat{\sigma}'}\right|$ in any dimension for the barycentric
dual edges, so that $(\Delta_{d}x)_{\sigma}$ reduces to a sum over
normals of a set of modified faces, which cannot be expected to close, in general. 

The second property is 
\begin{enumerate}
\item[6.] \emph{Positivity} of the weights, $w_{\sigma\sigma'}>0.$ It is also called Markov property \cite{Kigami:2001wk} and is directly related to Osterwalder--Schrader positivity. The latter is crucial for a Euclidean quantum field theory to yield unitarity in the corresponding Lorentzian theory after Wick rotation \cite{Osterwalder:1973tq}. 
\end{enumerate}
Positivity holds if all the volumes in the weights are positive. This is generally true for barycentric duals. For circumcentric duals, the situation is less general. Positivity does hold for circumcentric duals of regular complexes (where the circumcenters lie in the simplices).

However, this is not the case for irregular circumcentric duals. When a circumcenter does not lie inside the simplex, the part of the dual length associated with this simplex is negative such that, in some cases the sum of the two parts is negative (see figure \ref{fig2}), inducing negative Laplace matrix weights. 

Therefore, we see that, as anticipated, the choice of geometry of the dual complex is
crucial, yielding different properties for the discrete Laplacian. In quantum gravity, in particular in the investigation of its possible fractal structure, the barycentric dual is to be preferred.

Indeed, in this context, the null condition, symmetry and positivity are generally required. They are even taken as {\it the} defining properties in fractal spectral theory \cite{Kigami:2001wk} (see \ref{appA}). In contrast, it could be expected on general grounds that standard locality and linear precision might be violated. Although we do have locality for $\Delta_{d}$ in our simplicial context, the relation between such discrete and continuum notions of locality is not immediate. Indications of a breakdown of standard locality actually exist in several approaches to quantum gravity (e.g., \cite{CES}). Also, in fractional calculus, which can be used as an effective description of fractal and other anomalous spacetimes, the Laplacian may be composed by fractional integro-differential operators, which are non-local (by the dependence on non-neighboring points) \cite{frc4,frc1,frc2,fra6}.

Linear precision is not needed either because we are not
in flat space and its only relevance is as an asymptotic property in
the continuum limit to flat spaces. But as we have argued, this is already fulfilled
up to higher order corrections. That this works despite the
lack of linear precision can be easily understood by noting that the average
difference between circumcentric and barycentric dual lengths is only of
higher order in the scale of refined triangulations.
Thus, as far as quantum gravity is concerned, this seems enough since it does not seem reasonable to enforce properties of the continuum flat-space Laplacian exactly in the discrete theory.  
Also, fractional spacetimes are a continuum example where this property is violated, in all self-adjoint Laplacians (also in the second-order one, due to the presence of a measure weight to the right of the derivatives) \cite{frc4,frc3}.

As for why positivity should then be satisfied, instead, the reasons are the following. One is simply, in a sense, by exclusion, i.e., once we have decided that linear precision can be dropped, it makes sense to try to enforce as many as possible of the other properties. 
A second reason is that {\it all} quantum gravity approaches we consider are phrased as standard quantum theories on the lattice. 
The present discrete-calculus formalism is applicable only to their Euclidean versions, which we will discuss and which indeed are the ones best understood. Reflection positivity has not been yet directly related to unitarity in this context, still we expect such relation to exist, even if it is not realized by a simple Wick rotation. Moreover, it is not clear at all if there are mechanisms for recovering positivity in a continuum limit, if this is not enforced in the discrete operator. Therefore, it seems preferable to maintain it in the definition of the discrete theory. 

Third, one immediate application we have in mind for our Laplacian operator is the investigation of the geometric properties of states and histories in quantum gravity models, by means of the calculation of the spectral dimension. Indeed, this has been a major field of research in several discrete quantum gravity approaches, like dynamical triangulations \cite{AJL4,BeH} and tensor models \cite{GuR}, and, more recently, spin foams and LQG \cite{Mod08,CaM,MPM}. The calculation of the spectral dimension uses the discrete Laplacian operator for defining a test diffusion process taking place on the discrete structures defining quantum gravity states and histories. Positivity of such Laplacian is a necessary requirement in order to be able to have a properly defined diffusion process and thus a sensible spectral dimension observable. 


\section{Generalizations and applications}\label{sec3}


\subsection{Generalizations of simplicial pseudo-manifolds}

So far, we have detailed the formalism for primal simplicial pseudo-manifolds equipped with
a geometry. There are two possible generalizations which are important
for applications: pseudo-manifolds with boundary and more general
polyhedral complexes instead of simplicial ones. We sketch such generalizations without going into the details, as the construction is actually straightforward.

\subsubsection*{Boundaries.}

An abstract simplicial pseudo $d$-manifold $K$ is allowed to have a boundary $\partial K$ when the non-branching condition is relaxed. The $(d-1)$-simplices comprised in $\partial K$ have to be faces of only one $d$-simplex each. Therefore, the non-branching condition for simplicial pseudo-manifolds with a boundary states that each $(d-1)$-simplex is the face of one or two $d$-simplices. The other conditions of section \ref{hod} remain.

Then, $\partial K$, or more precisely all the elements of $(\partial K)_{d-1}$,
can be obtained from the action of the boundary operator $\partial$
on the $d$-chain comprising all $d$-simplices: 
\[
|\partial K\rangle=\partial\underset{\sigma_{d}\in K_{d}}{\sum}|\sigma_{d}\rangle\,,
\]
since the interior $\left(d-1\right)$-simplices cancel pairwise because of orientation.

The boundary $\partial K$ is just a $(d-1)$-subcomplex of $K$.
Without the original non-branching condition holding, the construction
of a dual $\star\partial K$ is only
possible using the simplicial subdivision explained above yielding
half-lines, or in general half-cells dual to face simplices. These
are distinguished as exterior cells $\hat{\sigma}^{e}\in\star\partial K$
from the usual interior ones $\hat{\sigma}^{i}=\hat{\sigma}\in\star K\backslash\star\partial K$.

For the calculus of fields $\phi\in\Omega^{p}(K)$, $p<d$, on the
simplicial pseudo-manifold nothing is changed besides exterior cells
having volumes accordingly. Only for $d$-forms $\phi\in\Omega^{d}(K)$,
it is necessary to define their boundary values extending their domain
from $K_{d}$ to $K_{d}\cup(\partial K)_{d-1}$. In general, it is desirable to have a boundary
field also for $d$-forms. One can then choose boundary conditions for such fields,
for example Dirichlet ones $\phi|_{\partial K}=\ast\phi|_{\star\partial K}=0$.

\subsubsection*{Cell complexes.}

Furthermore, one is interested in more general cell complex pseudo-manifolds on the primal side too. This poses no issue as far as cell complexes are concerned allowing for some simplicial decomposition for which one can use the formalism we have presented. Typically, one just wants to generalize from dual $(d+1)$-valent vertices to vertices of arbitrary valence, that is, from primal simplices to arbitrary polytopes. At the level of geometric realizations, the possibility of decompositions and hence the relation to simplicial pseudo-manifolds is obvious. One has only to take care of generalizing the definition  appropriately at the abstract combinatorial level.
Therefore, along the lines of \cite{Grady:2010wb}, we sketch how the
formalism is easily generalized to cell complexes obeying the three
pseudo-manifold conditions of section \ref{hod} at the topological level. 

A primal $p$-cell $\sigma_{p}$ now is a set of points homeomorphic
to a closed unit $p$-ball $B_{p}$; its boundary $\partial\sigma_{p}$
is the part of $\sigma_{p}$ homeomorphic to the boundary to $\partial B_{p}$.
It can be represented by the ordered set of vertices of a $p$-polytope. A cell $d$-complex $K$ is a collection of $p$-cells, $p=0,1,\dots ,d$, with the following two properties:
\begin{itemize}
\item The boundary $\partial\sigma_{p}$ of each $p$-cell $\sigma_{p}\in K$
is the union of some $\left(p-1\right)$-cells $\sigma_{p-1}\in K.$ 
\item The intersection of any two $p$-cells is empty or an element of the
boundary of both. 
\end{itemize}
As before, an orientation is given by the representation in terms of ordered
sets.

If $K$ is non-branching, strongly connected and dimensionally homogeneous,
it has a dual complex $\star K$ with cellular structure induced by
the adjacency relations of $K$, just as in the simplicial case. Then,
the whole formalism of discrete exterior calculus works through with
an appropriate definition of relative signs $\sgn(\sigma_{p-1},\sigma_{p})$.
All the formal definitions are already general enough to account
for this generalization.


\subsection{Applications: momentum transform and heat kernel}
\label{applications}

\subsubsection*{Momentum transform.}

Let us assume that the finite simplicial complex $K$ has a geometric
interpretation in terms of a set of finite, non-degenerate primal
and dual volumes. In particular, for the case of the Laplacian acting on a scalar function, the $d$-volumes $V_{\sigma_{d}}$ and dual edge lengths
$V_{\hat{\sigma}_{1}}$ should be non-vanishing. Then, the computation
of eigenvalues $\lambda$ and eigenfunctions $|\lambda\rangle$
of the Laplacian reduces to a purely linear algebraic issue. It depends
on the combinatorics of the simplicial complex as well as on the geometric
data. Note that in the defining equation
\[
\left(-\Delta_{d}e^{\lambda}\right)_{\sigma}:=-\left\langle \sigma|\Delta_{d}|\lambda\right\rangle =\lambda\left\langle \sigma|\lambda\right\rangle =:\lambda e_{\sigma}^{\lambda}
\]
indeed the asymmetric matrix elements $w_{\sigma\sigma'}/{V_{\sigma}}$
are essential. The eigenvalues $\lambda$ are defined with a relative
minus sign such that they are positive on, for example, closed pseudo-manifolds \cite{Chung:1997tk}.

If the matrix elements of the Laplacian are finite and well defined in the complex field, that is, if $\Delta$ is just a linear map in a finite vector space, then the Laplacian is diagonalizable and the eigenspaces of its eigenvectors comprise the vector space. Assuming this, the eigenfunctions $e_{\sigma}^{\lambda}$ of the Laplacian (where the label $\lambda$ is meant to run not only over eigenvalues but also over their multiplicities) form a complete orthonormal basis defining momentum space. The measure $V_{\lambda}$ of this space is thus induced by the norm chosen for the orthogonal eigenspace basis elements $|\lambda\rangle$ such that orthonormality, 
\[
\left\langle \lambda|\lambda'\right\rangle =\underset{\sigma}{\sum}V_{\sigma}^{(d)}e_{\sigma}^{\lambda}e_{\sigma}^{\lambda'*}=\frac{1}{V_{\lambda}}\delta_{\lambda\lambda'}\,,
\]
and consistently completeness, 
\[
\underset{\lambda}{\sum}V_{\lambda}|\lambda\rangle\langle\lambda|=\mathbbm{1}\,,
\]
hold.

For a momentum measure of the usual physical (energy) dimension $[V_{\lambda}]=d$, one could, for example, normalize the coefficients $e_{\sigma}^{\lambda}$ with respect to
the standard Euclidean measure. Although this choice of dimension is
not necessary, since any physical quantity will be automatically normalized
by the measure factors $V_{\lambda}$, it is the usual convention
in continuum physics to have position and momentum space measures of
reciprocal dimension. In this case, the momentum transform is an automorphism (see \cite{frc3,frc4} for further discussion).
Transformations of fields $\phi$ from position to momentum space and back are straightforwardly given by the resolution of the identity in either position or momentum space:\begin{eqnarray}
\phi^{\lambda} &=& \langle\phi|\underset{\sigma_{p}}{\sum}V_{\sigma_{p}}^{(d)}|\sigma_{p}\rangle\langle\sigma_{p}|\lambda\rangle=\underset{\sigma_{p}}{\sum}V_{\sigma_{p}}^{(d)}e_{\sigma}^{\lambda*}\phi_{\sigma}\,,\\
\phi_{\sigma}&=&\langle\phi|\underset{\lambda}{\sum} V_{\lambda}|\lambda\rangle\langle\lambda|\sigma\rangle=\underset{\lambda}{\sum} V_{\lambda} e_{\sigma}^{\lambda}\phi^{\lambda}\,.
\end{eqnarray}

\subsubsection*{Heat kernel.}

With a transform between position and momentum space at hand, one can easily deal also with functions of the Laplacian. We illustrate this with the example of the heat kernel, the solution to the heat equation on $K$ in terms of a continuous diffusion parameter $\tau$.

The formal expression of the heat kernel $e^{\tau\Delta_{d}}$ now
has a well-defined meaning on a simplicial pseudo-manifold $K$ for functions
on the dual complex: 
\begin{eqnarray}
K_{\sigma\sigma'}(\tau) & :=\left\langle \sigma'|e^{\tau\Delta_{d}}|\sigma\right\rangle 
=\langle\sigma'|e^{-\lambda\tau}\underset{\lambda}{\sum}V_{\lambda}|\lambda\rangle\langle\lambda|\sigma\rangle 
=\underset{\lambda}{\sum}V_{\lambda}e^{-\lambda\tau}e_{\sigma'}^{\lambda*}e_{\sigma}^{\lambda}\,.
\end{eqnarray}
We can use the heat kernel to calculate the diffusion of some initial matter distribution $\rho$ parametrized by $\tau$ to be
\begin{eqnarray}
\rho_{\sigma}(\tau) :=\langle\rho|K(\tau)|\sigma\rangle &=\langle\rho|e^{\tau\Delta_{d}}|\sigma\rangle
=\langle\rho|\underset{\sigma'}{\sum}V_{\sigma'}|\sigma'\rangle\langle\sigma'|e^{\tau\Delta_{d}}|\sigma\rangle\\
&=\underset{\sigma'}{\sum V_{\sigma'}}K_{\sigma\sigma'}(\tau)\langle\rho|\sigma'\rangle\nonumber
=\underset{\sigma'}{\sum}V_{\sigma'}K_{\sigma\sigma'}(\tau)\rho_{\sigma'}\,.
\end{eqnarray}
In particular, the heat kernel itself is the evolution
$\rho_{\sigma}(\tau)=K_{\sigma\sigma'}(\tau)$ for an initial distribution
$\rho_{\sigma}=\frac{1}{V_{\sigma}}\delta_{\sigma\sigma'}$ concentrated
on one simplex $\sigma'$. In the continuum, this initial condition would correspond to a diffusing test particle.

The trace per unit volume of the heat kernel, which gives the return probablity in diffusion
processes, becomes
\begin{eqnarray}
{\cal P}(\tau) & :=\mbox{tr} K_{\sigma\sigma'}(\tau)=\frac{1}{V}\underset{\sigma}{\sum}V_{\sigma}\underset{\lambda}{\sum}V_{\lambda}e^{-\tau\lambda}e_{\sigma}^{\lambda*}e_{\sigma}^{\lambda}
 =\frac{1}{V}\underset{\lambda}{\sum}V_{\lambda}e^{-\tau\lambda}\underset{\sigma}{\sum}V_{\sigma}e_{\sigma}^{\lambda*}e_{\sigma}^{\lambda}\nonumber\\
 & =\frac{1}{V}\underset{\lambda}{\sum}e^{-\tau\lambda}.
\end{eqnarray}

While the spectrum of the Laplacian gives a closed expression
for ${\cal P}(\tau)$ in many cases, for numerical computations
of combinatorially very large complexes it can alternatively be treated
as a random walk. In this case, local probabilities are given for jumping from one simplex
$\sigma$ to a neighbor $\sigma'$ given by the matrix elements ${w_{\sigma\sigma'}}/{V_{\sigma}}$
of the Laplacian. This is the technique used in dynamical triangulations
\cite{AJL4,BeH}, which will be discussed below, and random combs and multi-graphs  \cite{DJW1,AGW,GWZ1,GWZ2}.


\section{Classical expressions of the Laplacian}
\label{sec4}

The general form of the discrete Laplacian depends both on the combinatorial structure of the underlying simplicial complex and on its discrete geometry through the various volume factors. $\Delta$ takes then different concrete expressions, depending on the variables used to encode the geometry of the simplicial complex. These expressions would be needed for explicit calculations in different formulations of classical discrete gravity and, successively, in applications to quantum gravity models. In the following, we provide some examples for the discrete Laplacian constructed in the geometric variables used in various approaches to classical and quantum gravity.

\subsection{Regge edge-length variables}

The most common variables to describe the geometry of a simplicial
pseudo-manifold are the edge lengths $\left\{ l_{ij}\right\} $. In the standard
version of Regge calculus \cite{Regge:1961ct,Loll:1998ue}, these are
taken as configuration space for the geometries of piecewise flat triangulations.

The expressions for primal volumes are well known in the Regge
literature, so the only geometric data needed for defining the dual scalar
Laplacian $\Delta_{d}$ are the dual edge lengths $\hat{l}_{\sigma\sigma'}$.
We subdivide the dual edges into two parts $\hat{l}^{\sigma}$ and $\hat{l}^{\sigma'}$,
 associated, respectively, with the simplices $\sigma$ and $\sigma'$, so that $\hat{l}_{\sigma\sigma'}=\hat{l}^{\sigma}+\hat{l}^{\sigma'}$. These dual edge lengths depend on the chosen embedding of dual complex into the primal one.
 
In the barycentric case, when $\hat{l}_{\hat{i}}^{\sigma}$ is the length of the edge dual to the face $\sigma_{d-1}=(012\dots \hat{i}\dots d)$ contained inside the simplex $\sigma_{d}=(012\dots d)$, it is given by (see \ref{sec:Geometric-Data})
\[
\hat{l}_{\hat{i}}^{\sigma}=\frac{1}{d\left(d+1\right)}\sqrt{d\underset{j}{\sum}l_{ij}^{2}-\underset{(jk)}{\sum}l_{jk}^{2}}\,.
\]
Then, the matrix elements of the Laplacian (equation \Eq{lapelements}) for $\sigma\cap\sigma'=\left(012\dots d\right)\cap\left(0'12\dots d\right)=\left(12\dots d\right)$
have the form
\[
\frac{w_{\sigma\sigma'}}{V_{\sigma}}=d\left(d+1\right)\frac{1}{V_{012\dots d}}\frac{V_{12\dots d}}{\hat{l}_{\hat{0}}^{\sigma}+\hat{l}_{\hat{0}'}^{\sigma'}}\,.
\label{BaryEdgeLap}
\]
These are well defined on simplicial geometries satisfying the
strong generalized triangular inequalities, that is, $V_{\sigma_{p}}>0$
for all $0<p\le d$. In particular, these conditions ensure that the
dual lengths $\hat{l}_{i}^{\sigma}$ are non-zero and positive.

This is not the case for the circumcentric dual where each $\hat{l}_{\hat{0}}^{\sigma}\in\R$
can be negative or vanishing, and thus it is possible to have $\hat{l}_{\hat{0}}^{\sigma}+\hat{l}_{\hat{0}'}^{\sigma'}=0$. This pole in the expression for the Laplacian, moreover, cannot be absorbed into the volumes as they depend only on the
edges of $\sigma$ but not of $\sigma'$. On the other hand, except
for these singularities, the circumcentric Laplacian might be well defined
even on degenerate geometries with $V_{\sigma_{d}}=0$. This is true, for example,
for $d=2,3$, where explicit expressions of the circumradius are known (again \ref{sec:Geometric-Data}).
In $d=2$,
\[
\frac{w_{\left(ijk\right)\left(jkl\right)}}{A_{ijk}}=\frac{8}{\pm\left(l_{ij}^{2}+l_{ik}^{2}-l_{jk}^{2}\right)\pm\frac{A_{ijk}}{A_{jkl}}\left(l_{jl}^{2}+l_{kl}^{2}-l_{jk}^{2}\right)}\,,
\label{CircEdgeLap2d}
\]
and in $d=3$ 
\begin{eqnarray}
\fl \frac{w_{(ijkl)(ijkm)}}{V_{ijkl}}=12A_{ijk}^{2}
&&\Bigg[\pm\sqrt{\left(2A_{ijk}\mathcal{A}_{ijkl}\right)^{2}-\left(3l_{ij}l_{jk}l_{ki}V_{ijkl}\right)^{2}} \nonumber\\
&&\pm\frac{V_{ijkl}}{V_{ijkm}}\sqrt{\left(2A_{ijk}\mathcal{A}_{ijkm}\right)^{2}-\left(3l_{ij}l_{jk}l_{ki}V_{ijkm}\right)^{2}}\Bigg]^{-1}.
\label{CircEdgeLap3d}
\end{eqnarray}
The sign of each dual length part $\hat{l}^{\sigma}$ is positive if the circumcenter lies inside the $d$-simplex $\sigma$ and negative if outside. With these descriptions of the Laplacian at hand, one can compare with other discrete Laplacians in the literature.

\subsubsection*{Sorkin's discrete Laplacian.}

In \cite{Sorkin:1975kv}, a formalism with special `barycentric'
coordinates (not to be confused with the mathematical notion, where
unit vectors are attached to corners) is developed. As done also in \cite{Dittrich:2012uj},
it can be expressed in terms of the dihedral angles as a `cotangens' Laplacian
(with an inverse-volume factor) for primal scalar fields. In $d=2$,
with
$\alpha_{ij}^{\sigma_{2}}$ the angle opposite to the edge
$(ij)$ in the triangle $\sigma_{2}$, it is given by
\[
-(\Delta_{0}\phi)_{i}=\frac{1}{V_{\star(i)}}\underset{j}{\sum}\left(\underset{\sigma_{2}\ni(ij)}{\sum}\cot\alpha_{ij}^{\sigma_{2}}\right)(\phi_{i}-\phi_{j})\,,
\]
and it is easy to show its equivalence to the Laplacian coming from discrete calculus with  circumcentric duals. (Elementary geometric arguments yield $\hat{l}_{\hat{i}}^{\left(ijk\right)}=\sqrt{R^{2}-\left({l_{jk}}/{2}\right)^{2}}=({l_{jk}}/{2})\cot\alpha_{ij}^{\sigma_{2}}$.) In $d=3$,
\[
-(\Delta_{0}\phi)_{i}=\frac{1}{V_{\star(i)}}\underset{j}{\sum}\left(\underset{\sigma_{3}\ni(ij)}{\sum}l_{\hat{i}\hat{j}}^{\sigma_{3}}\cot\alpha_{ij}^{\sigma_{3}}\right)(\phi_{i}-\phi_{j})\,,
\]
where the opposite dihedral angle $\alpha_{ij}^{\sigma_{2}}$
now is between faces sharing the opposite edge $l_{\hat{i}\hat{j}}^{\sigma_{3}}$
in the tetrahedron $\sigma_{3}$ \cite{haus}. From the equivalence in $d=2$, it is tempting to conjecture equivalence
also for $d\geq 3$, but this remains to be proven.

\subsubsection*{Laplacian in dynamical triangulations.}

A different way of encoding the simplicial geometry of a piecewise flat triangulation, still based on the Regge calculus description,
is to fix all edge lengths to some constant value, and allow only changes in the combinatorics of the simplicial complex itself.
This idea underlies the quantum gravity program of dynamical triangulations \cite{Loll:1998ue,Ambjorn:2011wg}.
For such equilateral configurations, the Laplacian coming from discrete calculus drastically simplifies
(up to an overall factor) to a purely combinatorial graph Laplacian \cite{Chung:1997tk} of the form \Eq{DAeq}:
\[
\Delta_d\propto D-A\,,
\]
where the weights here are $w_{\sigma\sigma'}=1$ if $\sigma$ and $\sigma'$ are adjacent.

While in the Lorentzian version, named causal dynamical triangulations,
this should be modified by introducing negative length squares for
time-like edges, this modification is not implemented since the theory is Wick rotated to Euclidean signature and actual calculations are performed in a reduced ensemble of Euclidean triangulations (those that can indeed be obtained by Wick rotating Lorentzian ones) \cite{Ambjorn:2011wg}.

\subsection{First-order Regge calculus with \texorpdfstring{$(d-1)$}{}-face variables}

An alternative version to edge-length Regge calculus is in terms of
the $(d-1)$-face normals $\omega^{\sigma_{d-1}}(\alpha)$ (expressed in the reference frame of the $d$-simplex $\sigma^{\alpha}$) and Lorentz rotations (parallel transports) $U(\alpha,\alpha')$ from frame to frame across neighboring simplices. In turn, the latter define holonomies (around closed plaquettes) 
$W_{\alpha}(h)=U_{\alpha,\alpha+1}U_{\alpha+1,\alpha+2}\dots U_{\alpha-1,\alpha}$, which are rotations in the plane orthogonal to hinges $h\in K_{d-2}$
\cite{Caselle:1989cd,Barrett:1999ba,Gionti:2005gi} and measure the local curvature. The class angles corresponding to the holonomies
are therefore the deficit angles $\theta_{h}=2\pi-\sum_\alpha\theta_{h}^{\alpha}$, as could be
obtained from the dihedral angles $\theta_{h}^{\alpha}$ at the hinge
$h$ in each $d$-simplex $\sigma^{\alpha}$ sharing it (see also \cite{Hamber:2009zz}).

We show how all geometric data needed for the Laplacian $\Delta_{d}$
have an expression in terms of the face normals $\omega^{\sigma_{d-1}}(\alpha)$. 

While the $(d-1)$-volumes are just the modulus of the face normals themselves, 
\[
V_{\sigma_{d-1}}=\left|\omega^{\sigma_{d-1}}(\alpha)\right|\,,
\]
the $d$-volumes of simplices $\sigma^{\alpha}$ can also be expressed
by $d$ of the face normals $\omega^{i}(\alpha)=\omega^{\sigma_{d-1}=(012\dots \hat\imath\dots d)}$
as \cite{Caselle:1989cd} 
\[
V_{\alpha}\equiv V_{\sigma^{\alpha}}=\left[\frac{1}{d!}\epsilon^{I_{1}\dots I_{d}}\epsilon_{i_{1}\dots i_{d}j}\omega_{I_{1}}^{i_{1}}(\alpha)\dots \omega_{I_{d}}^{i_{d}}(\alpha)\right]^{\frac{1}{d-1}},
\]
where capital indices $I, J,\dots$ are in internal space. 
By the closure relations, it does not matter which face $(012\dots \hat\jmath\dots d)$
is left out if $\sigma^{\alpha}$ is closed. Alternatively, one could also
average over the choices of reference face.

An explicit expression of dual lengths can only be obtained using
position coordinates on $\sigma^{\alpha}$ as functions of the face
normals. Barycentric coordinates $z(\alpha)$, that is, coordinates
for which the sum over vertices satisfies $\sum_{i=1}^{d+1}z_{i}^{I}(\alpha)=0$,
can be derived inverting the expression of the face normals in terms
of discrete vielbeins (see equation (\ref{eq:Vielbein})) in these coordinates
\cite{Caselle:1989cd}:
\[
\omega_{I}^{i}(\alpha)=\frac{1}{\left(d-1\right)!^{2}}\underset{k\ne i}{\sum}\epsilon_{J_{1}\dots J_{d-1}I}\epsilon^{i,i_{1}\dots i_{d-1},k}z_{i_{1}}^{J_{1}}(\alpha)\dots z_{i_{d-1}}^{J_{d-1}}(\alpha)\,,
\]
leading to 
\[
z_{i}^{I}(\alpha)=\frac{1}{\left(d-1\right)!}\frac{1}{\left(V_{\alpha}\right)^{d-2}}\underset{k\ne i}{\sum}\epsilon^{J_{1}\dots J_{d-1}I}\epsilon_{i,i_{1}\dots i_{d-1},k}\omega_{J_{1}}^{i_{1}}(\alpha)\dots \omega_{J_{d-1}}^{i_{d-1}}(\alpha)\,.
\]
The barycentric dual length is particularly simple in these coordinates.
It is just the distance from the barycenter of the tetrahedron with
coordinate $z^{I}=0$  to the barycenter of a face
\[
\hat{l}_{\hat{i}}^{\sigma}=\left|\underset{j\ne i}{\overset{}{\sum}}z_{j}[\omega^{i}(\sigma)]\right|\,.
\]
For the circumcentric case, no such simplification can be expected.
Still, primal edge lengths can be expressed in the coordinates $z(\alpha)$, taking then advantage of the above expressions (equations \Eq{BaryEdgeLap}, \Eq{CircEdgeLap2d} and \Eq{CircEdgeLap3d}).

As an example, we can give the (further simplified) expressions in $d=3$.
On $\sigma^{\alpha}=(ijkl)$ (suppressing the frame label $\alpha$),
\[
\fl z_{i}^{I}=\frac{1}{2}\frac{1}{V_{\alpha}}\underset{r\ne i}{\sum}\epsilon^{IJK}\epsilon_{imnr}\omega_{J}^{m}\omega_{K}^{n}=\frac{1}{2V_{\alpha}}\left(\omega^{j}\times\omega^{k}+\omega^{k}\times\omega^{l}+\omega^{j} \times\omega^{l}\right)^{I},
\]
and the tetrahedron volume in terms of three of its face triangles is
\[
\left(V_{\alpha}\right)^{2}=\frac{1}{6}\epsilon^{IJK}\epsilon_{ijkl}\omega_{I}^{i}\omega_{J}^{j}\omega_{K}^{k}\,.
\]
Therefore, the dual length is 
\begin{eqnarray}
\fl \hat{l}_{i}^{\alpha} &= \frac{1}{3}\left|z_{j}+z_{k}+z_{l}\right|
=\frac{1}{6V_{\alpha}}\left|\omega^{j}\times\omega^{k}+\omega^{k}\times\omega^{l}+\omega^{l}\times\omega^{j}\right|\nonumber\\
\fl &=\frac{\sqrt{\underset{(mn)\in(jkl)}{\sum}\left[\omega_{m}^{2}\omega_{n}^{2}-(\omega_{m}\cdot\omega_{n})^{2}+(\omega_{m}\cd\omega_{r})(\omega_{r}\cdot\omega_{n})-(\omega_{m}\cdot\omega_{n})\omega_{r}^{2}\right]}}{6V_{\alpha}}\,.
\end{eqnarray}
Using the closure condition $\sum\omega^{i}=0$, this further simplifies to 
\[
\hat{l}_{i}^{\alpha}=\frac{1}{2V_{\alpha}}\left|\omega^{j}\times\omega^{k}\right|=\sqrt{\omega_{j}^{2}\omega_{k}^{2}-\left(\omega_{j}\cdot\omega_{k}\right)^{2}}\label{baryfacevar}
\]
for some faces $j,k$. The matrix elements \Eq{lapelements} of the Laplacian $\Delta_{d}$
can then easily be computed combining all the above expressions.

Finally, we note that the volume form $\omega^{h}(\alpha)$ of a hinge
$h=\sigma_{d-2}$ can be expressed in terms of two normals to two faces $\sigma^{\alpha+1,\alpha}$,
$\sigma^{\alpha,\alpha+1}$ sharing it, in the frame of $\sigma^{\alpha}$ \cite{Caselle:1989cd}: 
\[
\omega_{IJ}^{h}(\alpha)=\frac{1}{V_{\alpha}}\omega_{[I}^{\alpha-1,\alpha}(\alpha)\omega_{J]}^{\alpha,\alpha+1}(\alpha)\,,
\]
where square brackets denote anti-symmetrization of the indices. This gives a connection to flux variables, discussed in the next section, which are exactly these $(d-2)$-face normals. 


\subsection{Flux and area-angle variables}

In $d=4$, a useful alternative set of variables in simplicial geometry
are the bivectors $b_{ijk}^{IJ}=e_{ij}^{I}\wedge e_{ik}^{J}$ associated
with triangles $(ijk)$ (or their internal Hodge duals $X_{ijk}^{IJ}=\epsilon_{\ KL}^{IJ}b_{ijk}^{KL}$), 
known as fluxes, and playing a prominent role in both canonical loop quantum gravity and spin-foam models \cite{Rovelli:2004wb, Baratin:2010ti, Baratin:2011hc}. In a geometric
4-simplex $(ijklm)$, the triangle areas are 
\[
A_{ijk}=|X_{ijk}|\,,
\]
and volumes of tetrahedra can be computed using three of the fluxes associated with the four triangles on their boundary \cite{Barrett:1998fp}, regarding the bivectors as linear maps: 
\[
V_{ijkl}^{2}=\frac{8}{9}\mbox{Tr}\left(\ast X_{ijk}\left[\ast X_{jkl},\ast X_{kli}\right]\right)\,.
\]
Volumes of 4-simplices can be taken from the wedge product of two
fluxes not lying in the same $3$-hyperplane (thus not belonging to the same tetrahedron):
\[
V_{ijklm}=|X_{ijk}\wedge X_{ilm}|\,.
\]
Primal edge lengths can be expressed using the generalized sine formula as 
\[
l_{ij}^2=2\frac{|X_{ijk}|^2|X_{ijl}|^2-(X_{ijk}\cdot X_{ijl})^2}{\mbox{Tr}\left(\ast X_{ijk}\left[\ast X_{jkl},\ast X_{kli}\right]\right)}\,.
\]
This gives all the buildings blocks for explicit expressions  (equations \Eq{BaryEdgeLap}, \Eq{CircEdgeLap2d} and \Eq{CircEdgeLap3d}) of the
barycentric and circumcentric discrete Laplacian $\Delta_{d}$ with elements \Eq{lapelements}.

In the spin representation in $d=3+1$ LQG and $d=4$ spin foams (adapted to a simplicial context), the easiest variables to use are triangle areas and $3$-volumes of tetrahedra. However, it is known that they form an overcomplete set of data to specify a four-dimensional simplicial geometry and should be supplemented by additional constraints whose explicit form is not known \cite{Barrett:1999fa,Makela:1994hm}. A more natural choice is to use areas $A_{ijk}$ and dihedral
angles $\phi_{k,l}^{ij}$ between faces $(ijk)$ and $(ijl)$ hinged
at the common edge $(ij)$ \cite{Dittrich:2008hg}. This set of data encodes the same
information as the fluxes $X_{ijk}$. In these variables, the relevant geometric data to compute the discrete Laplacian have the following expressions.  The $3$-volumes are

\[
V_{ijkl}^{2}=\frac{A_{ijk}}{9}\sqrt{\underset{j}{\sum}A_{ijl}^{2}\sin^{2}\phi_{k,l}^{ij}A_{jkl}^{2}\sin^{2}\phi_{i,l}^{jk}-\underset{(ij)}{\sum}A_{ijl}^{4}\sin^{4}\phi_{k,l}^{ij}}\,,
\]
from which the $4$-volumes are obtained via the generalized sine
law\footnote{The angles $\theta_{l,m}^{ijk}$ between 3-simplices $(ijkl)$ and
$(ijkm)$ are functions of the area dihedral angles according to \cite{Dittrich:2008hg}
\begin{equation}\nonumber
\cos\theta_{l,m}^{ijk}=\frac{\cos\phi_{k,l}^{ij}-\sin\phi_{l,m}^{ij}\sin\phi_{m,k}^{ij}}{\cos\phi_{l,m}^{ij}\cos\phi_{m,k}^{ij}}\,.
\end{equation}}
\[
V_{ijklm}=\frac{3}{4}\frac{1}{A_{ijk}}V_{ijkl}V_{ijkm}\sin\theta_{l,m}^{ijk}[\phi]\,,
\]
as well as the primal edge lengths
\[
l_{ij}=\frac{2}{3}\frac{1}{V_{ijkl}}A_{ijk}A_{ijl}\sin\phi_{k,l}^{ij}\,.
\]
Again, this is all the information needed to build the Laplacian $\Delta_{d}$.


\section{Laplacian in models of quantum geometry}\label{sec5}

With the classical expressions of the dual scalar Laplacian $\Delta_{d}$
in the appropriate geometric variables at hand, one can take one's favorite
model of quantum gravity and promote $\Delta_{d}$ to a quantum observable. For instance, one can either take $\Delta_{d}$ as an operator acting on quantum states of spatial
geometries in a canonical theory (e.g., in an LQG context) or as a classical function to be
path integrated over with the quantum measure of a covariant theory (within a spin-foam or simplicial path integral setting). We now discuss briefly how such calculations could be set up, leaving explicit computations for future study.

In both types of approaches, the main challenge beyond a purely
formal quantization is to deal with possible singularities of the
matrix entries of the Laplacian, coming from
the inverse $d$-volumes in the barycentric case and from the inverse
dual length part in the circumcentric case. 
In a canonical setting, these singularities may prevent the definition of the Laplacian operator as a bounded operator; in the covariant setting, they may produce divergences in explicit evaluations. Obviously, whether or not such difficulties arise depend on the details of the quantum theory considered, and depending on the precise structure of the Hilbert space of states or the path-integral measure, as well as on the exact
classical expression to be quantized, they may not necessarily pose
a problem.

Furthermore, for many purposes, it is not the Laplacian $\Delta_{d}$
as such but its functions $f[\Delta_{d}]$ which are of interest.
These need not have the same quantization issues (e.g., possible singularities) as the Laplacian
itself.

A good example is the trace of the heat kernel $\mathcal{P}(\tau)$,
discussed in the classical simplicial setting above (section \ref{applications}). Since it is of
the general form $\mathcal{P}(\tau)\sim e^{\tau\Delta}$, one would expect that it
vanishes exactly in those cases where the Laplacian is singular (see the example
in \ref{sec:Simple-Example:-Degenerate}). Thus, one may even envisage cases in which
observable functions of the Laplacian $f[\Delta_{d}]$, inserted within quantum geometric evaluations (e.g., path integrals), might even help to suppress pathological configurations corresponding to degenerate or divergent geometries.

In the context of a quantum theory of pure geometry without
any dynamical matter, there are reasons to believe that, quite in general,
only global functions of the Laplacian are suitable complete observables  (beyond the kinematical level) since
they are invariant under diffeomorphisms. The heat trace is a good
example of an observable meeting these conditions. 


\subsection{Laplacian in canonical formalism}

The best developed canonical approach to quantum gravity
is LQG \cite{Rovelli:2004wb,Thiemann:1111397}. The kinematical Hilbert
space of states of spatial geometry is defined as a projective limit
of Hilbert spaces $\mathcal{H}_{\Gamma}$ of states associated with
graphs $\Gamma$. Under certain assumptions \cite{Bombelli:2009hg},
they can be considered as the 1-skeleton $\Gamma=(\star K)_{1}$ of
the dual of a combinatorial pseudo-manifold $K$. Since the valency of the
nodes in $\Gamma$ is left arbitrary in LQG, the complex has to be
polyhedral in general, though often one restricts to the lowest non-trivial
(non-vanishing volume) valency of $d+1$, corresponding to primal simplicial
pseudo-manifolds. (In principle, one can take an expression of the Laplacian obtained
from the geometric interpretation in a pseudo-manifold setting and apply
it even to graphs $\Gamma$ which are not in the skeleton of the dual to a pseudo-manifold,
as long as all the variables are defined.)

The LQG states are cylindrical functions
$\psi_{\Gamma}(h_{\hat{\sigma}_{1}})$ of holonomies of the gauge
group $G=SU(2)$ on the links $\hat{\sigma}_{1}$ of the graphs $\Gamma$. These variables encode the extrinsic geometry of the spatial slice.
The same states can be transformed into functions of representations $j_{\hat{\sigma}_{1}}$
on the links and intertwiners $i_{\hat{\sigma}_{0}}$ between them
on the nodes $\hat{\sigma}_{0}$, called spin network states $\psi_{\Gamma}(j_{\hat{\sigma}_{1}},i_{\hat{\sigma}_{0}})$.
A further possibility is to transform into a basis of fluxes $X_{\hat{\sigma}_{1}}$
on the links, valued in the Lie algebra of the group \cite{Baratin:2011hc}. These sets of dual variables encode the intrinsic geometry of the spatial slice.

The spin network states are the eigenstates of a commuting set of
local geometric observables. In $d=2+1$, these are the primal edge
length operators $\widehat{l_{\sigma_{1}}}$ dual to graph links $\hat{\sigma}_{1}=\star\sigma_{1}$,
with squared spectrum proportional to the Casimir of the group $G=SO(3)\cong SU(2)$:
\[
\widehat{l_{\sigma_{1}}^{2}}\psi_{\Gamma}(j_{\hat{\sigma}_{1}},i_{\hat{\sigma}_{0}})\sim\left[j_{\star\sigma_{1}}(j_{\star\sigma_{1}}+1)+c\right]\psi_{\Gamma}(j_{\hat{\sigma}_{1}},i_{\hat{\sigma}_{0}})\,,
\]
with $c={\rm const}$ being a quantization ambiguity.

In $d=3+1$, the same holds with the difference that it is now primal triangles (more generally, polygons)
 to be dual to the graph
links $\hat{\sigma}_{1}=\star\sigma_{2}$, and the spins are then their areas $\widehat{A_{\hat{\sigma}_{2}}}$ for which 
\[
\widehat{A_{\sigma_{2}}^{2}}\psi_{\Gamma}(j_{\hat{\sigma}_{1}},i_{\hat{\sigma}_{0}})\sim\left[j_{\star\sigma_{2}}(j_{\star\sigma_{2}}+1)+c\right]\psi_{\Gamma}(j_{\hat{\sigma}_{1}},i_{\hat{\sigma}_{0}})\,.
\]
The 3-volume operator 
$\widehat{V_{\sigma_{3}}}$ for the tetrahedron (more generally, 3-cell) dual to a graph vertex has a (more complicated)
spectrum in terms of the intertwiners $i_{\star\sigma_{3}}$ \cite{Thiemann:1111397}.

Concerning length operators $\widehat{l_{\sigma_{1}}}$ for primal edges, there are several definitions available in the literature. In one such definition \cite{Bianchi:2008ib}, eigenstates of $\widehat{l_{\sigma_{1}}}$
are linear combinations of the intertwiners and the operators
$\widehat{l_{\sigma_{1}}}$ corresponding to edges of the primal 3-cell neither commute with the volume operator of the same 3-cell $\widehat{V_{\sigma_{3}}}$ nor with one another in the case of intersecting edges.

A natural way to promote the spatial Laplacian to a quantum operator
would therefore be to regard it as a function of these basic geometric
observables. In $d=2+1$, on states with simplicial combinatorics,
that is, 3-valent graphs $\Gamma$, this is fairly straightforward
as the commuting set of length operators captures the whole simplicial
geometry. Thus, the two-dimensional spatial Laplacian $\Delta_{2}$ can be formally quantized as a composition of length operators:
\[
\widehat{\Delta_{2}}=\Delta_{2}[\widehat{l_{\sigma_{1}}}]\,.
\]
In practice, to avoid the issue of zeros in the denominator in either
the barycentric or circumcentric description, a regularization\footnote{In \cite{Bianchi:2008ib}, for example, a Tikhonov regularization \cite{Tikhonov:1977ug}
is used to cure inverse-volume issues.} or linearization of the classical expression $\Delta_{2}[l_{\sigma_{1}}]$ (equation \Eq{BaryEdgeLap} or \Eq{CircEdgeLap2d})
is needed to achieve a well-defined operator $\Delta_{2}[\widehat{l_{\sigma_{1}}}]$.

In $d=3+1$, this quantization cannot work as easily because
the commuting set of operators $\widehat{A_{\sigma_{2}}^{2}}$ and
$\widehat{V_{\sigma_{3}}}$ is not sufficient to determine a simplicial spatial 
geometry. Therefore, the quantum Laplacian $\widehat{\Delta_{3}}$
can only be expressed as a function of operators at least a pair of
which is non-commuting. A consequence is that $\widehat{\Delta_{3}}$
cannot be diagonalized in the spatial geometry states on a given graph.
This fact is less problematic than it may look at first sight. Ultimately,
pure states of quantum geometry cannot be expected to have a geometry
in a classical, e.g., simplicial sense. Only semi-classical coherent
states peaked on a classical geometry are supposed to have this
meaning. On such states, it should be possible to obtain a well-defined
action and expectation value of $\widehat{\Delta_{3}}$.

Since we do have expressions of $\Delta_{3}$ in three dimensions
in terms of face normals (equation \Eq{baryfacevar} and so on), that is, fluxes in the canonical setting,
appropriate types of coherent states to be used are those in flux variables studied in
\cite{Oriti:2012kx,Oriti:2012hl}. As $\widehat{\Delta_{3}}$ is now built from non-commuting operators,
there are also ordering ambiguities, and the same issue of regularization of possible inverse volume divergences will also have to be dealt with.

\subsubsection*{Comparison with other proposed Laplacians in the LQG context.}

We will close this subsection discussing briefly our Laplacian with other proposals appeared in the LQG literature, usually defined in the context of matter Hamiltonians. These proposals are indeed different from ours.

From the Hamiltonian of a non-relativistic point particle on an LQG
space, one can read off the following Laplacian
$\widehat{\Delta_{3}}$ \cite{Rovelli:2010ic}. Assuming that the Hamiltonian is diagonal
in the Hilbert space of spin network states $|s\rangle$, the result
of a discretization procedure is (in the notation of \cite{Rovelli:2010ic})
\begin{equation}
\widehat{\Delta_{3}}\sim\underset{s,l\in s}{\sum}A_{l}^{2}| s,\underset{\sim}{l}\rangle\langle s,\underset{\sim}{l}|
\end{equation}
on a position basis of the particle on the links of the graph $|\underset{\sim}{l}\rangle$,
where the underlining with a tilde indicates that an inverse-volume
factor is included in the definition of this basis.

In the position basis of dual points $\hat{\sigma}_{0}$ (dual to primal simplices) natural for
the dual scalar function, its expectation value on a spin network state $|s\rangle$ is
\begin{equation}
\langle s|\widehat{\Delta_{3}}|s\rangle\sim\underset{\sigma}{\sum}\underset{\sigma'\sim\sigma}{\sum}\frac{A_{\sigma\sigma'}^{2}}{V_{\sigma}^{2}}|\sigma\rangle\langle\sigma|-\underset{\sigma}{\sum}\underset{\sigma'\sim\sigma}{\sum}\frac{A_{\sigma\sigma'}^{2}}{V_{\sigma}V_{\sigma'}}|\sigma\rangle\langle\sigma'|\,,\label{eq:RoveVido}
\end{equation}
where $A_{\sigma\sigma'}=V_{\sigma\cap\sigma'}$ are the areas of the primal faces
dual to the links connecting $\sigma$ and $\sigma'$.

Obviously, this differs from the Laplacian operator \Eq{DAeq} coming from discrete calculus. If
the inverse volumes are understood to belong to the position states,
the above expression is just a graph Laplace matrix with weights $A_{\sigma\sigma'}^{2}$. This is the definition used in \cite{Rovelli:2010ic}.  On the other hand, for the Laplacian to have the right dimension, the volumes would have to be considered as part of its definition (and not hidden in the position basis) and the weights are then, as in the formula above, ${A_{\sigma\sigma'}^{2}}/(V_{\sigma}V_{\sigma'})$. The advantage of the first choice of position basis and Laplacian with exclusive dependence on the areas, for an application to LQG, is that this Laplacian only needs, for its evaluation, the geometric information that is present in pure spin network states, bypassing the issues discussed in the previous section. On the other hand, one might then question whether this choice captures the whole geometric content of the Laplacian, as the one coming from discrete calculus does, and gives an operators with the right properties. Our analysis would suggest that this is not the case, but the above simpler operator could nevertheless represent a useful approximation in some contexts.

A Laplacian of a similar type was also considered in \cite{Mod08} in the context of an evaluation of the spectral dimension in LQG and spin foams. More precisely, the scaling of the Laplacian was all that was needed in that setting, and it was taken to be given just by the area spectrum, so that in practice it amounted as dealing with a diagonal Laplacian.

Another LQG Laplacian appears in \cite{Thiemann:1998hn}, within the Hamiltonian for a scalar field. In order to deal with the issue of inverse volumes, one uses Thiemann's trick of substituting inverse 3-volumes with Poisson brackets of holonomies and (powers of) 3-volumes. The Laplacian operator then takes the form 
\[\label{thila}
\left(\widehat{\Delta_3}\phi\right)_{\hat{\sigma}_{0}}\sim\frac{N(\hat{\sigma}_{0})}{E(\hat{\sigma}_{0})^{2}}\underset{\underset{v_{\Delta}=\hat{\sigma}_{0}}{\Delta}}{\sum}\mbox{tr}(\hat h[\hat h^{-1},\hat{V}^{\frac{3}{4}}])^{4}(\phi_{s(\Delta)}-\phi_{\hat{\sigma}_{0}})\,,
\]
where $N(\hat{\sigma}_{0})$ and $E(\hat{\sigma}_{0})$ are some combinatorial
factors depending on the vertex $\hat{\sigma}_{0}$, $\hat h$ is the holonomy
operator and the sum effectively runs over neighbors too. The precise
structure, in particular of the spectrum, is not known, so a more detailed comparison with the Laplacian coming from discrete calculus, from which it clearly differs, is not possible. 


\subsection{Laplacian in covariant models}

In covariant theories of quantum gravity, the Laplacian lives in spacetime itself rather than on spatial slices only. Even for the spatial Laplacian in LQG, a covariant counterpart in terms
of a spin-foam model might be necessary to evaluate it within a physical
scalar product.

Such covariant approaches are typically formulated as discretized path integrals.
The sum over 4-geometries for a given boundary 3-geometry is defined
for geometries on a simplicial pseudo-manifold $K$
(e.g., in Regge calculus), or on its dual complex
(e.g., in spin-foam models, which can also be re-expressed as simplicial gravity path integrals), and may include a sum over these complexes as well
(dynamical triangulations and group field theories). We discuss briefly the templates for the evaluation of the discrete Laplacian as a geometric observable in these contexts.

\subsubsection*{Quantum Regge calculus.}

The formalism of discrete calculus is most easily applied to the Regge approach. This is, first of all, because Regge calculus works directly on a simplicial pseudo-manifold $K$. Second, because the configuration space summed over consists only of simplicial geometries, even in the quantum version. In the latter, this condition has to be imposed by special constraints, namely, the strict generalized triangle inequalities. These demand the volumes of all $p$-simplices to be positive, $V_{\sigma_{p}}>0$. On such simplicial geometries, the discrete Laplacian is automatically well defined (no issues with degeneracies or singularities). Neither in edge-length variables $l_{ij}^{2}$ nor in the $(d-1)$-face normal variables $\omega$ there are any problems in expressing the Laplacian as a classical observable in Regge calculus. 

In principle, one could therefore go straight to the quantum theory on a given triangulation $|K|$ in the path-integral formulation, given an appropriate measure $\mu_{\rm Regge}^{|K|}=[\mathcal{D}l_{ij}]\,e^{iS_{\rm Regge}[l_{ij}]}$ or $\mu_{\rm Regge}^{|K|}=[\mathcal{D}U_{\alpha\beta}][\mathcal{D}\omega_{\alpha\beta}]\,e^{iS_{\rm Regge}[U_{\alpha\beta},\omega_{\alpha\beta}]}$, and consider the quantum expectation value 
\[
\left\langle f[\Delta_{d}]\right\rangle ^{|K|}=\int\mu_{\rm Regge}^{|K|}f[\Delta_{d}]\,.
\]
While the Regge action $S_{\rm Regge}$ is well known in both cases, the
definition of the exact measure of such a model of quantum gravity is still a pending challenge, with respect to the imposition of the generalized triangle inequalities as well as the issue of simplicial symmetries \cite{Hamber:2009zz, Dittrich:2008pw}.

\subsubsection*{Spin foams and related path integrals.}

This path-integral expectation value can be considered also in spin foams, an approach generalizing the concept of Regge geometries \cite{Dupuis:2012ub} where a precise form of the measure
can be motivated from a discretization of the Holst--Plebanski action.

By the motivation of spin foams as a path-integral version of LQG
defined via spatial graphs $\Gamma$, that is, 1-complexes, the discrete
counterpart of spacetime is usually defined as a 2-complex $\mathcal{C}$ (hence
the name `foam'). Analogously to the discussion of the canonical
case, in a strict sense the discrete Laplacian $\Delta_{d}$ is therefore
applicable only to 2-complexes being the 2-skeleton of a dual $d$-complex,
$\mathcal{C}=(\star K)_{2}$. In spin foams, only $\left(d+1\right)$-valent
vertices are considered such that the primal complex $K$ would be
indeed a simplicial complex. Nevertheless, an expression of the simplicial
$\Delta_{d}$ could still be generalized to the setting of arbitrary
2-complexes $\mathcal{C}$, as long as they are equipped with enough
geometric data for all the volumes in $\Delta_{d}$ to be defined.

Just as in LQG, the geometry variables could be fluxes, holonomies,
or their spin representations of the full gauge group $G$ on the
edges $\hat{\sigma}_{1}\in\mathcal{C}$. A spin foam in the strict
sense of the name refers to the latter. Most generally, it is defined
as a path-integral state sum over representations $j_{\hat{\sigma}_{2}}$
and intertwiners $i_{\hat{\sigma}_{1}}$ by a measure factorizing
into amplitudes $\mathcal{A}_{\sigma_{p}}$ on faces, edges and vertices
on $\mathcal{C}$ \cite{Perez:2012uz}: 
\begin{eqnarray}
Z^{\mathcal{C}} & =\underset{\{j_{\hat{\sigma}_{2}}\},\{i_{\hat{\sigma}_{1}}\}}{\sum}\mu_{\rm SF}^{\mathcal{C}}\nonumber\\
 & =\underset{\{j_{\hat{\sigma}_{2}}\},\{i_{\hat{\sigma}_{1}}\}}{\sum}\underset{\hat{\sigma}_{2}\in\mathcal{C}}{\prod}\mathcal{A}_{\hat{\sigma}_{2}}(j_{\hat{\sigma}_{2}})\underset{\hat{\sigma}_{1}\in\mathcal{C}}{\prod}\mathcal{A}_{\hat{\sigma}_{1}}(j_{\hat{\sigma}_{2}},i_{\hat{\sigma}_{1}})\underset{\hat{\sigma}_{0}\in\mathcal{C}}{\prod}\mathcal{A}_{\hat{\sigma}_{0}}(j_{\hat{\sigma}_{2}},i_{\hat{\sigma}_{1}})\,.
\end{eqnarray}
On the other hand, this is just the spin-foam representation of an
underlying more general path integral which could equally well be
expressed in holonomies $g$ or fluxes $X$ with corresponding measures:
\[
Z^{\mathcal{C}}=\int\left[{\cal D} g_{\hat{\sigma}_{2}}\right]\mu_{g}^{\mathcal{C}}=\int\left[{\cal D} X_{\hat{\sigma}_{2}}\right]\mu_{X}^{\mathcal{C}}\,.
\]

Since these variables are directly related to the LQG variables in
the canonical theory, the discussion of the possibility to express the Laplacian
through them is similiar. Particularly simple is the $d=3$ case of the so-called Ponzano--Regge model with a measure $\mu_{\rm PR}^{\mathcal{C}}$ defined in terms of
the dimension of representations associated with edges of the dual complex and of $6j$-symbols associated
with vertices \cite{Ponzano:1968wi}. From the length operator in
LQG, an interpretation of primal lengths dual to the foam faces $\hat{\sigma}_{2}$
can be induced such that 
\[
l_{\sigma_{1}}^{2}=l_{\hat{\sigma}_{2}}^{2}=j_{\hat{\sigma}_{2}}(j_{\hat{\sigma}_{2}}+1)+c\,.
\]
This defines $\Delta_{3}=\Delta_{3}(j_{\hat{\sigma}_{2}})$ on $\mathcal{C}$
in its edge-length version (equation \Eq{BaryEdgeLap} or equation \Eq{CircEdgeLap3d}) and one has a formal expectation value of functions of the Laplacian:
\[
\left\langle f[\Delta_{3}]\right\rangle _{\rm PR}^{\mathcal{C}}=\underset{\{j_{\hat{\sigma}_{2}}\}}{\sum}\mu_{\rm PR}^{\mathcal{C}}f[\Delta_{3}(j_{\hat{\sigma}_{2}})]\,.
\]
It is then straightforward generalizing to the well-understood case with LQG spin network
states $|s\rangle$ on the boundary of $\mathcal{C}$, where the state sum
is running only over internal labels with fixed boundary configurations
induced from $|s\rangle$ \cite{Noui:2005js}.

As already noted, the geometric interpretation of spin-foam configurations
is more general than Regge geometries. While the trivial intertwiners
$i_{\hat{\sigma}_{1}}=i_{\star\sigma_{2}}$ implicit in the $6j$-symbols
constrain the primal triangles $\sigma_{2}$ to close, there are
no conditions for the tetrahedra $\sigma_{3}$ (more generally, top-dimensional simplices) to close too. Therefore,
the volumes $V_{\sigma_{3}}(l_{\sigma_{1}}^{2})$ might take complex
values or even vanish. Since they appear in the denominator of the Laplace matrix elements (equation \Eq{BaryEdgeLap}), this may result in poles of the Laplacian.

In $d=4$, it is more challenging to obtain a version of $\Delta_{4}$ in terms of the spin representation labels via the LQG-induced relation to primal areas $A_{\sigma_{2}}$
(dual to foam faces $\hat{\sigma}_{2}$) and $3$-volumes $V_{\sigma_{3}}$
(dual to foam edges $\hat{\sigma}_{1}$). While the number of labels is in principle large enough, the issue of configurations not uniquely specifying a simplicial geometry discussed above becomes relevant again. A convenient set of variables in which to compute the expectation value of $\Delta_{4}$ is obtained in the flux representation of the
state sum, which then takes the form of a $BF$-like simplicial path integral. The fluxes $X_{\hat{\sigma}_{2}}$ are now the volume forms of primary faces $\sigma_{2}$, which can also be equivalently associated with their dual faces
$\hat{\sigma}_{2}=\star\sigma_{2}$ in the foam.

A general remark is the following. In any discrete path integral, whether configurations on which $\Delta_{d}$ is divergent lead to divergences of the overall sum over quantum geometric configurations or not depends very much on the dynamics encoded in the measure. 
If there were divergences, they could be treated with an appropriate regulator or, when possible, by directly excluding the singular configurations from the path integral. On the other hand, many spin-foam amplitudes are generically divergent even before inserting other geometric observables, and some regularization/renormalization might be needed from the start, anyway. Proper calculations of Laplacian-based observables would have to be then phrased in this regularized context. We will do this in future work. 


\section{Conclusions and outlook}\label{sec6}

We have employed discrete calculus, known from computational science \cite{Desbrun:2005ug,Grady:2010wb}, as a formalism for differential operators and arbitrary fields at a fundamentally discrete level, more precisely on simplicial complexes and their combinatorial dual complexes. This should open up novel ways to investigate the physical and geometric properties of simplicial theories of quantum gravity.

With respect to \cite{Desbrun:2005ug}, we chose a different, more physical convention where a geometric space measure is explicitly taken into account. The formalism was presented in a convenient bra-ket notation and, therefore, slightly generalized, thus providing a setting to rigorously define a discrete Laplacian operator $\Delta$. We analyzed the action of the Laplacian on scalar fields living on vertices of the dual complex. The discrete Laplacian can be required to satisfy several properties, coming from continuum properties, from usual lattice gauge theory, from the fractal literature or from reasonable physical requirements. Whether these properties are satisfied or not by the discrete Laplacian we considered depends on the precise geometric embedding of the dual complex into the primal one. In particular, we have shown that the barycentric version may be preferred to the circumcentric one because it does lead to a positivity property that is the discrete counterpart of Osterwalder--Schrader positivity.

The formalism can be made sufficiently general to be extended to polyhedral complexes and complexes with a boundary. Also, the Laplacian enters the definition of an invertible momentum transform to a representation of fields on its eigenspaces. This generalization of the Fourier transform works on arbitrary discrete geometries and can be effectively used to handle functions of the Laplacian such as the heat kernel and, from that, the spectral dimension of spacetime. The latter will be the subject of a companion paper \cite{COT2}. The use of (functions of) the discrete Laplacian as a geometric observable to unravel the geometry of quantum gravity states and histories is indeed one application we envisage for our results. Another application is as a necessary ingredient for defining matter coupling in discrete models of quantum gravity.

These results are ready to be applied to various gravity approaches. We gave explicit expressions of the Laplacian in geometric variables used in loop quantum gravity, spin foams, Regge calculus and dynamical triangulations: edge lengths, face normals, fluxes and area-angle variables. We discussed how to apply these expressions to specific models, either in a canonical or covariant formalism, and the issues to be tackled. Fluxes seem to be the type of variables with the most general applicability, i.e., for combinatorics other than those of $d$-complexes, as they can be used to define general polyhedral geometries. Operator issues about inverse volumes and dual lengths (present inside $\Delta$) could be cured in various ways: in canonical theory, by regularization or linearizations, in the covariant one by regularization, renormalization precedures or appropriate modifications of models. At any rate, we also noted that considering functions of the Laplacian, rather than the Laplacian itself, may make these issues irrelevant for practical purposes, as discussed in the example of the heat kernel and in \cite{COT2}. In particular, the spectral dimension in LQG and spin-foam models can be computed and is well defined. 

We conclude with a comment on the continuum limit. In continuum flat space, the discrete Laplacian \Eq{eq:ScalarLaplace} reduces to the second-order continuum Laplace operator:
\begin{equation}\label{coli}
\Delta \to \sum_{\mu=1}^d\partial_\mu^2\,.
\end{equation}
However, the limit to the continuum in a discrete quantum gravity model is much less trivial because it must include quantum dynamics, a wealth of geometric information (curvature, effective measures respecting quantum symmetries and so on) and physical matter fields. Thus, the correct physical description of a quantum geometry in a large-scale/low-energy/semi-classical regime may remain elusive in several interesting cases. 

The diffusion equation is a crystalline example in this respect. In a discrete setting, it is defined via a test field $\phi$  obeying $(\partial_\tau-\Delta)\phi=0$, with some initial condition $\phi|_{\tau=0}$ in the abstract diffusion time and where curvature effects are ignored \cite{SVW2,frc4,fra6}. In the naive sense of equation \Eq{coli}, this expression reduces to the continuum diffusion equation in flat Euclidean space, with the consequence that the spectral dimension of the continuum manifold $\mathbb{R}^d$ on which the diffusion process takes place is the classical one $d$. However, if one first computes the effective spectral dimension in a genuinely discrete (and quantum) setting (such as causal dynamical triangulations, for instance \cite{AJL4}, or spin foams \cite{Mod08,CaM,MPM,COT2}), and taking into account the full quantum dynamics, the output would differ from $d$ at any given scale, even in semi-classical or continuum approximations and even in the zero-curvature limit. $\mathbb{R}^d$ is not necessarily the effective manifold ${\cal M}_{\rm cont}$ representing the physical continuum limit of the quantum-fluctuating geometry in the large-scale regime. As briefly discussed in section \ref{lapr}, the physical continuum limit is a black-box procedure which can also generate effective continuous Laplacians (in the sense of the operator governing diffusion processes) which may violate one or more of the properties of the discrete $\Delta$, and of the standard continuum one, including locality and the effective order of the operator \cite{frc4,fra6}. 

The task of getting control over this important aspect of quantum gravity models goes beyond the scope of the present work. Yet, the stage in which this issue can be tackled in the near future has been hopefully improved by the results presented here.


\ack{GC and DO acknowledge financial support through a Sofia Kovalevskaja Award at the AEI. JT acknowledges financial support by the Andrea von Braun Foundation and, at an earlier stage of the work, by the FAZ Foundation.}


\appendix

\section{Laplacians on fractals}\label{appA}

In the spectral theory of deterministic fractals, Laplacians are defined as a limit of Laplace
matrices $\Delta^{\Gamma_{m}}$ on a sequence of graphs $\Gamma^{m}$
approximating the fractal which can then be identified with $\underset{m\ra\infty}{\lim}\Gamma^{m}$. In general \cite{Kigami:2001wk}, one defines Laplacians $-\Delta^{\Gamma_{m}}$ on the vertex set of the graphs $\Gamma^{m}$ as symmetric linear operators with three properties: positive definiteness, the null condition
and the Markov property. These are exactly the conditions satisfied
by the barycentric version of the discrete Laplacian presented in section \ref{lapr} since
positive definiteness follows from symmetry and positivity \cite{Wardetzky:2008kk}.

The graph Laplacian $\Delta^{\Gamma_{m}}$ needs two modifications to define
the Laplacian on the fractal: a `renormalization factor' $r^{-m}$
according to the graph approximation and a volume factor $V_{\hat{\sigma}_{0}}^{(m)}$
for the evaluation of a function at a point on the graph $\hat{\sigma}_{0}\in\Gamma^{m}$,
similar to discrete calculus. The volume factor depends on the self-similar
(space) measure $\mu$ on the fractal since 
\begin{eqnarray}
\langle \hat{\sigma}_{0}|\Delta^{\Gamma_{m}}|\phi\rangle  & =\int\d\mu\ \chi_{\hat{\sigma}_{0}}\Delta^{\Gamma_{m}}\phi=\int\d\mu\ \psi_{\hat{\sigma}_{0}}^{(m)}\Delta^{\Gamma_{m}}\phi\nonumber\\
 & \approx\left(\int\d\mu\ \psi_{\hat{\sigma}_{0}}^{(m)}\right)\left(\Delta^{\Gamma_{m}}\phi\right)_{\hat{\sigma}_{0}}=:V_{\hat{\sigma}_{0}}^{(m)}\left(\Delta^{\Gamma_{m}}\phi\right)_{\hat{\sigma}_{0}}.
\end{eqnarray}
Here, the characteristic function $\chi_{\hat{\sigma}_{0}}$ for $\hat{\sigma}_{0}$
on the fractal is approximated by the so-called harmonic splines $\psi_{\hat{\sigma}_{0}}^{(m)}$,
which are functions sufficiently peaked on $\hat{\sigma}_{0}$ on
the fractal and identical to the Dirac distribution on the graphs
$\Gamma^{m}$ \cite{Strichartz:2006tm}.

Eventually, the Laplacian on the fractal is defined as the limit
\[
\left(\Delta\phi\right)_{\hat{\sigma}_{0}}:=\underset{m\ra\infty}{\lim}\frac{1}{V_{\hat{\sigma}_{0}}^{(m)}r^{m}}\langle \hat{\sigma}_{0}|\Delta^{\Gamma_{m}}|\phi\rangle\,.
\]
While the volume factor in the known and understood examples of deterministic fractals
is just an overall constant independent of the approximation level
$m$, the exact renormalization factor is crucial to obtain a neither
vanishing nor trivial Laplacian \cite{Strichartz:2006tm}. 


\section{Geometric data of simplices\label{sec:Geometric-Data}}

For a simplicial complex $K$ with a geometric realization as a piecewise
linear space, the frame field can be considered as a set of discrete
edge vectors 
\begin{equation}
e^{I}=e_{\mu}^{I}\d x^{\mu}\quad\mapsto\quad e_{ij}^{I}(\alpha)=\left[x_{i}(\alpha)-x_{j}(\alpha)\right]^{I}\,,\label{eq:Vielbein}
\end{equation}
where the coordinates $x(\alpha)$ are given by a choice of origin
and frame for every $d$-simplex
$\sigma^{\alpha}$ the edge $(ij)$ is face of. The index $\alpha=1,2,\dots ,N_{d}$ labels the $d$-simplices. Accordingly, the volume form of a $p$-simplex $\sigma_{p}$ in the
coordinates of a $d$-simplex of which it is a face is
\[
\omega_{I_{p+1}\dots I_{d}}^{\sigma_{p}}(\alpha)=\epsilon_{I_{1}\dots I_{d}}\underset{k=1}{\overset{p}{\prod}}e_{k}^{I_{k}}(\alpha)\,,
\]
in terms of $p$ linear independent edge vectors $e_{k}$ belonging
to $\sigma_{p}$. The $p$-volume $\sigma_{p}$ is the norm of the
volume form
\[
V_{\sigma_{p}}=\left|\omega^{\sigma_{p}}\right|=\frac{1}{p!}\sqrt{\underset{I_{p+1}<\dots <I_{d}}{\sum}\left|\omega_{I_{p+1}\dots I_{d}}^{\sigma_{p}}\right|^{2}}\,.
\]


\subsection{Edge-length variables \label{sub:Edge-length-variables}}

In the edge-length variables $\{l_{ij}^{2}\}$, the primal (simplicial) volumes can be
obtained from the Cayley--Menger determinant 
\[
V_{\sigma_{p}}=\frac{1}{p!}\frac{\left(-1\right)^{\frac{p+1}{2}}}{2^{\frac{p}{2}}}\left|\begin{array}{cccc}
0 & 1 & \cdots & 1\\
1 & 0 & l_{ij}^{2} & \cdots\\
\vdots & l_{ij}^{2} & \ddots\\
1 & \vdots &  & 0
\end{array}\right|^{\frac{1}{2}}.
\]
In particular,
\[
V_{\sigma_{2}}=\frac{1}{4}\sqrt{\underset{i}{\sum}\left(2l_{ij}^{2}l_{ik}^{2}-l_{jk}^{4}\right)}
\]
and, after some manipulations, 
\begin{eqnarray}
\fl V_{\sigma_{3}} &=& \frac{1}{12}\sqrt{\underset{(ij)}{\sum}l_{ij}^{2}\left(l_{ik}^{2}l_{jl}^{2}+l_{il}^{2}l_{jk}^{2}-l_{ij}^{2}l_{kl}^{2}\right)-\underset{(ijk)}{\sum}l_{ij}^{2}l_{ik}^{2}l_{jk}^{2}}\,,\\
\fl V_{\sigma_{4}} &=& \frac{1}{96}\left[\underset{(ij)(kl)}{\sum}l_{ij}^{4}l_{kl}^{4}+\underset{(ij)(k)}{\sum}\left(l_{li}^{2}l_{ik}^{2}l_{kj}^{2}l_{jm}^{2}+l_{mi}^{2}l_{ik}^{2}l_{kj}^{2}l_{jl}^{2}-l_{ij}^{4}l_{kl}^{2}l_{km}^{2}\right)\right.\nonumber\\
\fl && \qquad\left.-2\underset{(ijkl)}{\sum}l_{ij}^{2}l_{jk}^{2}l_{kl}^{2}l_{li}^{2}-4\underset{(ij)}{\sum}l_{ij}^{2}l_{kl}^{2}l_{lm}^{2}l_{mk}^{2}\right]^{\frac{1}{2}}\,,
\end{eqnarray}
where all sums run over all subsimplices of the given kind.


\subsection{Barycentric dual volumes}

The dual volumes are much more involved and we will consider
only dual lengths. In the barycentric case, the length $\hat{l}_{i}^{\sigma}$ of the
part in one simplex $\sigma_{d}=(012\dots d)$ of an edge dual to the
face $\sigma_{d-1}=(012\dots \hat{i}\dots d)$ is given by 
\[
\hat{l}_{i}^{\sigma}=\frac{1}{d\left(d+1\right)}\sqrt{d\underset{j}{\sum}l_{ij}^{2}-\underset{(jk)}{\sum}l_{jk}^{2}}\,.
\]
This can be seen as follows. In coordinates $x$, the position of the barycenter $x_{bc}$ of a $p$-simplex is 
\[
x_{bc}=\frac{1}{p+1}\underset{i=0}{\overset{p}{\sum}}x_{i}\,.
\]
The distance from the barycenter of $\sigma_{d}$ to the barycenter
of $\sigma_{d-1}=(12\dots d)$ in these coordinates then is 
\begin{equation}
\hat{l}_{0}^{\sigma}=\left|\frac{1}{d+1}\underset{i=0}{\overset{d}{\sum}}x_{i}-\frac{1}{d}\underset{i=1}{\overset{d}{\sum}}x_{i}\right|=\left|d\,x_{0}-\frac{1}{d(d+1)}\underset{i=1}{\overset{d}{\sum}}x_{i}\right|\,,\label{eq:BaryLength}
\end{equation}
and choosing coordinates where $x_{0}$ is the origin and using $x_{i}\cdot x_{j}=g_{ij}=\frac{1}{2}\left(l_{0i}^{2}+l_{0j}^{2}-l_{ij}^{2}\right)$
\cite{Hamber:2007wi}, this reduces to 
\begin{eqnarray}
\hat{l}_{0}^{\sigma} & =\frac{1}{d(d+1)}\sqrt{\left(\underset{i=1}{\overset{d}{\sum}}x_{i}\right)^{2}}=\frac{1}{d\left(d+1\right)}\sqrt{\underset{i}{\sum}x_{i}^{2}-2\underset{(ij)}{\sum}x_{i}\cdot x_{j}}\nonumber\\
 & =\frac{1}{d\left(d+1\right)}\sqrt{d\underset{i}{\sum}l_{0i}^{2}-\underset{(ij)}{\sum}l_{ij}^{2}}\,.
\end{eqnarray}

Besides the simple two dimensional case, this formula was also proven before for the tetrahedron \cite[theorem 187]{AltshillerCourt:1964vd}.


\subsection{Circumcentric dual volumes}

In the circumcentric case, in $d=2$ one gets dual edge lengths from
the circumradius $R_{ijk}=\frac{l_{ij}l_{ik}l_{jk}}{4A_{ijk}}$: 
\begin{eqnarray}
\hat{l}_{jk}^{(ijk)} & =\sqrt{R_{ijk}^{2}-\left(\frac{l_{jk}}{2}\right)^{2}}=\frac{l_{jk}}{2}\sqrt{\frac{l_{ij}^{2}l_{ik}^{2}}{4A_{ijk}^{2}}-1}\nonumber\\
 & =\frac{l_{jk}}{4A_{ijk}}\sqrt{l_{ij}^{4}+l_{jk}^{2}+l_{ki}^{2}-l_{ij}^{2}l_{ik}^{2}-2(l_{jk}^{2}l_{ij}^{2}+l_{jk}^{2}l_{ik}^{2})}\,.
\end{eqnarray}
Since 
\begin{eqnarray}
4l_{jk}^{2}l_{ik}^{2}-16A_{ijk}^{2} & =l_{ij}^{4}+l_{jk}^{4}+l_{ki}^{4}-2(l_{jk}^{2}l_{ij}^{2}+l_{jk}^{2}l_{ik}^{2}-l_{ij}^{2}l_{ik}^{2})\nonumber\\
 & =(l_{ij}^{2}+l_{ik}^{2}-l_{jk}^{2})^{2}\,,
\end{eqnarray}
this simplifies to 
\[
\hat{l}_{jk}^{(ijk)}=\frac{l_{ij}^{2}+l_{ik}^{2}-l_{jk}^{2}}{4A_{ijk}}\frac{l_{jk}}{2}\,.
\]
The matrix elements of the Laplacian are 
\begin{eqnarray}
\frac{w_{\left(ijk\right)\left(jkl\right)}}{A_{ijk}} & =\frac{1}{A_{ijk}}\frac{2}{\pm\sqrt{\frac{l_{ij}^{2}l_{ik}^{2}}{4A_{ijk}^{2}}-1}\pm\sqrt{\frac{l_{jl}^{2}l_{kl}^{2}}{4A_{jkl}^{2}}-1}}\\
& =\frac{4}{\pm\sqrt{l_{ij}^{2}l_{ik}^{2}-4A_{ijk}^{2}}\pm\frac{A_{ijk}}{A_{jkl}}\sqrt{l_{jl}^{2}l_{kl}^{2}-4A_{jkl}^{2}}}\nonumber\\
 & =\frac{8}{\pm\left(l_{ij}^{2}+l_{ik}^{2}-l_{jk}^{2}\right)\pm\frac{A_{ijk}}{A_{jkl}}\left(l_{jl}^{2}+l_{kl}^{2}-l_{jk}^{2}\right)}\,.
\end{eqnarray}
For $d=3$, there is a formula relating the circumradius $R$ of the
tetrahedron $(ijkl)$ to the area $\mathcal{A}_{ijkl}$ of a triangle
with the product of opposite edge lengths in the tetrahedron as its
edge lengths \cite{AltshillerCourt:1964vd}: 
\[
6V_{ijkl}R_{ijkl}=\mathcal{A}_{ijkl}\,.
\]
The circumcentric dual length to a face $(ijk)$ thus is 
\[
\hat{l}_{\hat{l}}^{(ijkl)}=\sqrt{R_{ijkl}^{2}-R_{ijk}^{2}}=\frac{\sqrt{\left(2A_{ijk}\mathcal{A}_{ijkl}\right)^{2}-\left(3l_{ij}l_{jk}l_{ki}V_{ijkl}\right)^{2}}}{12A_{ijk}V_{ijkl}}\,,
\]
and the Laplace weight 
\begin{eqnarray}
\fl w_{(ijkl)(ijkm)} &=& 12A_{ijk}^{2}\left[\pm\sqrt{\left(\frac{2A_{ijk}\mathcal{A}_{ijkl}}{V_{ijkl}}\right)^{2}-\left(3l_{ij}l_{jk}l_{ki}\right)^{2}}\right.\nonumber\\
\fl && \left.\pm\sqrt{\left(\frac{2A_{ijk}\mathcal{A}_{ijkm}}{V_{ijkm}}\right)^{2}-\left(3l_{ij}l_{jk}l_{ki}\right)^{2}}\right]^{-1}\,.
\end{eqnarray}
A simplification to avoid the square roots, as in $d=2$, remains to be found.


\section{Simple example: degenerate triangulation of the \texorpdfstring{$d$}{}-sphere \label{sec:Simple-Example:-Degenerate}}

To illustrate the formalism, we consider as an example the triangulation
of the $d$-sphere $S^{d}$ by two $d$-simplices labeled $a,b$ with
the same vertices $1,2,\dots ,d+1$ which are glued along all their faces
$(1,\dots ,\hat{i},\dots d+1)$. The weights of the Laplacian are $w_{(1,\dots ,\hat{i},\dots, d+1)}={V_{(1,\dots ,\hat{i},\dots, d+1)}}/{V_{\star(1,\dots ,\hat{i},\dots, d+1)}}$
but in the end only the degrees $D=D_{a}=D_{b}=\sum_{(1,\dots ,\hat{i},\dots, d+1)}w_{(1,\dots ,\hat{i},\dots, d+1)}$
enter in this specific example: 
\[
\left(-\Delta\phi\right)_{a}=\frac{1}{V_{a}}\underset{(1,\dots ,\hat{i},\dots d+1)}{\sum}w_{(1,\dots ,\hat{i},\dots, d+1)}\left(\phi_{a}-\phi_{b}\right)=\frac{D}{V_{a}}\left(\phi_{a}-\phi_{b}\right)\,,
\]
and for the simplex $b$ accordingly. The eigenvalues of the Laplacian
are $\lambda_0=0$ and $\lambda_1=V\,D/(V_{a}V_{b})$, which directly give
the heat trace and spectral dimension.

The eigenvectors $e_{\sigma}^{\lambda_0}=(1,1)/\sqrt{2}$ and $e_{\sigma}^{\lambda_1}=(V_{b},-V_{a})/\sqrt{2V_{a}V_{b}}$ are normed to a constant momentum basis measure of the inverse of the average volume
per simplex
\[
V_{\lambda_0}=V_{\lambda_1}=\frac{2}{V}\,, 
\]
and it is easily checked that they are orthogonal. Then, the heat kernel coefficients are 
\begin{eqnarray}
K_{\sigma\sigma'}(\tau) & =\frac{1}{V}\left(\begin{array}{cc}
V_{a}(1+\frac{V_{b}}{V_{a}}e^{-\frac{V\,D}{V_{a}V_{b}}\tau}) & V_{a}(1-e^{-\frac{V\,D}{V_{a}V_{b}}\tau})\\
V_{b}(1-e^{-\frac{V\,D}{V_{a}V_{b}}\tau}) & V_{b}(1+\frac{V_{a}}{V_{b}}e^{-\frac{V\,D}{V_{a}V_{b}}\tau})
\end{array}\right)\,,
\end{eqnarray}
and we can explicitly check that 
\[
K_{\sigma\sigma'}(\tau)\underset{\tau\ra\infty}{\ra}\left(\begin{array}{cc}
\frac{1}{V_{a}} & 0\\
0 & \frac{1}{V_{b}}
\end{array}\right)
\]
and that its trace is just 
\[
{\cal P}(\tau)=\frac{1}{V}\left(1+e^{-\frac{V\,D}{V_{a}V_{b}}\tau}\right)\,.
\]

For example, in $d=2$ using the edge-length variables $\{ l_{12},l_{13},l_{23} \}$ and the barycentric dual,
\[
D = \frac{3 l_{12}}{\sqrt{2 (l_{13}^2 + l_{23}^2)-l_{12}^2}}+\frac{3 l_{23}}{\sqrt{2 (l_{12}^2 + l_{13}^2) - l_{23}}}+\frac{3 l_{13}}{\sqrt{2 l_{12}^2 - l_{13}^2 + 2 l_{23}}}
\]
and the 2-volume is
\[
V_a = V_b = \frac{V}{2} = \frac{1}{4} \sqrt{2 l_{12}^2 (l_{23}^2 + l_{13}^2) - l_{12}^4 -(l_{23}^2 - l_{13}^2)^2}\,.
\]
In the equilateral case $l_{12} = l_{13} = l_{23} = l_\ast $, this trivializes to $D = 9$ and $V_a = V_b = \sqrt{3}/4$. 


\section*{References}

\providecommand{\href}[2]{#2}
\begingroup\raggedright
\endgroup


\begin{thebibliography}{99}

\bibitem{Rovelli:2004wb}
Rovelli C 2007 \emph{Quantum Gravity} (Cambridge: Cambridge University Press)
\bibitem{Thiemann:1111397}
Thiemann T 2007 \emph{Modern Canonical Quantum General Relativity} (Cambridge: Cambridge University Press)
\bibitem{Oriti:2001jh}  Oriti D {2001} \tia{Spacetime geometry from algebra: spin foam models for non-perturbative quantum gravity} \doin{10.1088/0034-4885/64/12/203}{Rep.\ Prog.\ Phys.}{}{64}{1489} (\arX{gr-qc/0106091}) 
\bibitem{Perez:2003wk}  P\'erez A {2003} \tia{Spin foam models for quantum gravity} 
\doin{10.1088/0264-9381/20/6/202}{Class.\ Quantum Grav.}{}{20}{R43} (\arX{gr-qc/0301113})
\bibitem{Perez:2012uz}
P{\'e}rez A {2013} \tia{The spin foam approach to quantum gravity}
\doin{10.12942/lrr-2013-3}{Living Rev.\ Rel.}{}{16}{3} 
\bibitem{GFT1} Oriti D {2009} \tia{Recent progress in group field theory} \doin{10.1063/1.3284386}{AIP Conf.\ Proc.}{}{1196}{209} (\arX{0912.2441})
\bibitem{GFT2} Oriti D 2012 \tia{The microscopic dynamics of quantum space as a group field theory}
\emph{Foundations of Space and Time: Reflections on Quantum Gravity} ed G Ellis \emph{et al} (Cambridge:
Cambridge University Press) (\arX{1110.5606})
\bibitem{Hamber:2009zz} Hamber H W 2008 \emph{Quantum Gravitation: The Feynman Path Integral Approach} (Amsterdam: Springer)
\bibitem{Ambjorn:2010kv} Ambj{\o}rn J, Jurkiewicz J and Loll R {2010} \tia{Quantum gravity as sum over spacetimes} \doin{10.1007/978-3-642-11897-5_2}{Lect.\ Notes\ Phys.}{}{807}{59} (\arX{0906.3947})
\bibitem{fra4}  Calcagni G {2011} \tia{Discrete to continuum transition in multifractal spacetimes}
 \doin{10.1103/PhysRevD.84.061501}{Phys.\ Rev.}{D}{84}{061501(R)} (\arX{1106.0295})
\bibitem{Niedermaier:2006up}
Niedermaier M and Reuter M {2005} \tia{The asymptotic safety scenario in quantum gravity}
  \doin{10.12942/lrr-2006-5}{Living Rev.\ Rel.}{}{9}{5}
\bibitem{Thiemann:1998hn}
Thiemann T {1998} \tia{Quantum spin dynamics (QSD): V. Quantum gravity as the natural regulator of the Hamiltonian constraint of matter quantum field theories}
  \doin{10.1088/0264-9381/15/5/012}{Class.\ Quantum Grav.}{}{15}{1281} (\arX{gr-qc/9705019})
\bibitem{Oriti:2002gy} Oriti D and Pfeiffer H {2002} \tia{Spin foam model for pure gauge theory coupled to quantum gravity}
\doin{10.1103/PhysRevD.66.124010}{Phys.\ Rev.}{D}{66}{124010} (\arX{gr-qc/0207041})
\bibitem{Speziale:2007ha} Speziale S {2007} \tia{Coupling gauge theory to spinfoam 3D quantum gravity}
\doin{10.1088/0264-9381/24/20/014}{Class.\ Quantum Grav.}{}{24}{5139} (\arX{0706.1534})
\bibitem{Oriti:2006kg} Oriti D and Ryan J {2006} \tia{Group field theory formulation of 3D quantum gravity coupled to matter fields} \doin{10.1088/0264-9381/23/22/027}{Class.\ Quantum Grav.}{}{23}{6543} (\arX{gr-qc/0602010})
\bibitem{Fairbairn:2007bu} Fairbairn W J and Livine E R {2007} \tia{3D spinfoam quantum gravity: matter as a phase of the group field theory} \doin{10.1088/0264-9381/24/20/021}{Class.\ Quantum Grav.}{}{24}{5277}
(\arX{gr-qc/0702125})
\bibitem{Bianchi:2010vy}
Bianchi E, Han M, Magliaro E, Perini C, Rovelli C and Wieland W 2010 \tia{Spinfoam fermions} \arX{1012.4719}
\bibitem{Han:2011uu}
Han M and Rovelli C 2011 \tia{Spinfoam fermions: PCT symmetry, Dirac
  determinant, and correlation functions} \arX{1101.3264}
\bibitem{Rovelli:2010ic}
Rovelli C and Vidotto F {2010} \tia{Single particle in quantum gravity and
  Braunstein--Ghosh--Severini entropy of a spin network}
  \doin{10.1103/PhysRevD.81.044038}{Phys.\ Rev.}{D}{81}{44038} (\arX{0905.2983})
\bibitem{AJL4}  Ambj{\o}rn J, Jurkiewicz J and Loll R {2005} \tia{The spectral dimension of the Universe is scale dependent} \doin{10.1103/PhysRevLett.95.171301}{Phys.\ Rev.\ Lett.}{}{95}{171301} (\arX{hep-th/0505113})
\bibitem{LaR5}  Lauscher O and Reuter M 2005 \tia{Fractal spacetime structure in asymptotically safe gravity} \doij{10.1088/1126-6708/2005/10/050}{J.\ High Energy Phys.}{}{JHEP10}{050}{2005} (\arX{hep-th/0508202})
\bibitem{Ben08} Benedetti D {2009} \tia{Fractal properties of quantum spacetime} \doin{10.1103/PhysRevLett.102.111303}{Phys.\ Rev.\ Lett.}{}{102}{111303} (\arX{0811.1396})
\bibitem{Mod08} Modesto L {2009} \tia{Fractal structure of loop quantum gravity} \doin{10.1088/0264-9381/26/24/242002}{Class.\ Quantum\ Grav.}{}{26}{242002} (\arX{0812.2214})
\bibitem{Hor3}  Ho\v{r}ava P {2009} \tia{Spectral dimension of the universe in quantum gravity at a Lifshitz point} \doin{10.1103/PhysRevLett.102.161301}{Phys.\ Rev.\ Lett.}{}{102}{161301} (\arX{0902.3657})
\bibitem{CaM}  Caravelli F and Modesto L 2009 \tia{Fractal dimension in 3D spin-foams} \arX{0905.2170}
\bibitem{Car09} Carlip S {2009} \tia{Spontaneous dimensional reduction in short-distance quantum gravity?} \doin{10.1063/1.3284402}{AIP Conf.\ Proc.}{}{1196}{72} (\arX{0909.3329})
\bibitem{BeH}   Benedetti D and Henson J {2009} \tia{Spectral geometry as a probe of quantum spacetime} \doin{10.1103/PhysRevD.80.124036}{Phys.\ Rev.}{D}{80}{124036} (\arX{0911.0401})
\bibitem{MPM}  Magliaro E, Perini C and Modesto L 2009 \tia{Fractal space-time from spin-foams} \arX{0911.0437}
\bibitem{Car10} Carlip S 2012 \tia{The small scale structure of spacetime}
 \emph{Foundations of Space and Time: Reflections on Quantum Gravity} ed G Ellis \emph{et al} (Cambridge: Cambridge University Press) (\arX{1009.1136})
\bibitem{SVW1}  Sotiriou T P, Visser M and Weinfurtner S {2011} \tia{Spectral dimension as a probe of the ultraviolet continuum regime of causal dynamical triangulations} \doin{10.1103/PhysRevLett.107.131303}{Phys.\ Rev.\ Lett.}{}{107}{131303} (\arX{1105.5646})
\bibitem{SVW2}  Sotiriou T P, Visser M and Weinfurtner S {2011}
  \tia{From dispersion relations to spectral dimension ---and back again}
  \doin{10.1103/PhysRevD.84.104018}{Phys.\ Rev.}{D}{84}{104018} (\arX{1105.6098})
\bibitem{AA}    Alesci E and Arzano M {2012} \tia{Anomalous dimension in semiclassical gravity} \doin{10.1016/j.physletb.2011.12.026}{Phys.\ Lett.}{B}{707}{272} (\arX{1108.1507})
\bibitem{frc4}  Calcagni G {2013} \tia{Diffusion in multiscale spacetimes} \doin{10.1103/PhysRevE.87.012123}{Phys.\ Rev.}{E}{ 87}{012123} (\arX{1205.5046})
\bibitem{RSnax} Reuter M and Saueressig F {2013} \doin{10.1007/978-3-642-33036-0_8}{Lect.\ Notes Phys.}{}{863}{185} (\arX{1205.5431})
\bibitem{COT2}  Calcagni G, Oriti D and Th\"urigen J (in preparation) preprint no.\ AEI-2013-196
\bibitem{Desbrun:2005ug}
Desbrun M, Hirani A N, Leok M and Marsden J E 2005 \tia{Discrete exterior calculus} \arX{math/0508341}
\bibitem{Hirani:2003ug}
Hirani A N 2003 Discrete exterior calculus \emph{PhD thesis} California Institute of Technology
\bibitem{Desbrun:2003db}
Desbrun M, Hirani A N and Marsden J E {2003} \tia{Discrete exterior calculus for variational problems in computer vision and graphics} \doin{10.1109/CDC.2003.1272393}{Proc.\ 42nd IEEE Conf.\ on Decision and Control}{}{5}{4902}
\bibitem{Gurau:2010iu} Gurau R {2010} \tia{Lost in translation: topological singularities in group field
  theory} \doin{10.1088/0264-9381/27/23/235023}{Class.\ Quantum\ Grav.}{}{27}{235023} (\arX{1006.0714})
\bibitem{Smerlak:2011ea} Smerlak M {2011} \tia{Comment on `Lost in translation: topological singularities in
  group field theory'} \doin{10.1088/0264-9381/28/17/178001}{Class.\ Quantum\ Grav.}{}{28}{178001}
  (\arX{1102.1844})
\bibitem{Gurau:2011hu} Gurau R {2011} \tia{Reply to comment on `Lost in translation: topological
  singularities in group field theory'}
 \doin{10.1088/0264-9381/28/17/178002}{Class.\ Quantum\ Grav.}{}{28}{178002} (\arX{1108.4966})
\bibitem{Weingarten:1977hy}
Weingarten D {1977} \tia{Geometric formulation of electrodynamics and general relativity in discrete space-time} \doin{10.1063/1.523124}{J.\ Math.\ Phys.}{}{18}{165}
\bibitem{Jourjine:1987iw}
Jourjine A N {1987} \tia{Discrete gravity without coordinates}
\doin{10.1103/PhysRevD.35.2983}{Phys.\ Rev.}{D}{35}{2983}
\bibitem{Itzykson:1983vc}
Itzykson C 1984 \tia{Fields on a random lattice} \href{http://www.springer.com/physics/book/978-0-306-41829-7}{\cob \emph{Progress in Gauge Field Theory}} ed G 't Hooft \emph{et al}
(New York: Plenum) 
\bibitem{Albeverio:1990ii} Albeverio S and Zegarlinski B {1990}
\tia{Construction of convergent simplicial approximations of quantum fields on Riemannian manifolds}
\doin{10.1007/BF02277999}{Commun.\ Math.\ Phys.}{}{132}{39}
\bibitem{Adams:1996ul} Adams D H 1996 \tia{$R$ torsion and linking numbers from simplicial abelian gauge theories}
\arX{hep-th/9612009}
\bibitem{Adams:1997iy} Adams D H {1997} \tia{A doubled discretisation of abelian Chern-Simons theory}
\doin{10.1103/PhysRevLett.78.4155}{Phys.\ Rev.\ Lett.}{}{78}{4155} (\arX{hep-th/9704150})
\bibitem{Sen:2000cr} Sen S, Sen S, Sexton J C and Adams D H {2000}
\tia{Geometric discretization scheme applied to the Abelian Chern--Simons theory}
\doin{10.1103/PhysRevE.61.3174}{Phys.\ Rev.}{E}{61}{3174} (\arX{hep-th/0001030})
\bibitem{Gross:2004vp} Gross P P W and Kotiuga P R 2004 \emph{Electromagnetic Theory and Computation: A Topological Approach} (Cambridge: Cambridge University Press)
\bibitem{Teixeira:2013ee} Teixeira F L {2013}
\tia{Differential forms in lattice field theories: an overview}
\doin{10.1155/2013/487270}{ISRN\ Math.\ Phys.}{}{2013}{487270}
\bibitem{Grady:2010wb} Grady L J and Polimeni J R 2010 \emph{Discrete Calculus: Applied Analysis on Graphs for Computational Science} (Dordrecht: Springer)

\bibitem{AlgebraicQFT} Haag R 2008 \emph{Local Quantum Physics: Fields, Particles, Algebras} (Berlin: Springer)
\bibitem{Dirac:1939ck} Dirac P A M {1939} \tia{A new notation for quantum mechanics}
\doin{10.1017/S0305004100021162}{Math.\ Proc.\ Camb.\ Phil.\ Soc.}{}{35}{416}
\bibitem{Nakahara:2003vq} Nakahara M 2003 \emph{Geometry, Topology, and Physics} (London: Taylor and Francis)
\bibitem{Rosenberg:1997to}
Rosenberg S 1997 \emph{The Laplacian on a Riemannian Manifold} (Cambridge: Cambridge University Press)
\bibitem{Kozlov:2008wc} Kozlov D 2008 \emph{Combinatorial Algebraic Topology} (Berlin: Springer)
\bibitem{Christ:1982kr} Christ N H, Friedberg R and Lee T D {1982} \tia{Random lattice field theory: general formulation}  \doin{10.1016/0550-3213(82)90222-X}{Nucl.\ Phys.}{B}{202}{89}
\bibitem{Christ:1982hv} Christ N H, Friedberg R and Lee T D {1982} \tia{Weights of links and plaquettes in a random lattice}   \doin{10.1016/0550-3213(82)90124-9}{Nucl.\ Phys.}{B}{210}{337}
\bibitem{Christ:1982bn} Christ N H, Friedberg R and Lee T D {1982} \tia{Gauge theory on a random lattice}
  \doin{10.1016/0550-3213(82)90123-7}{Nucl.\ Phys.}{B}{210}{310}
\bibitem{Mattiussi:1997jp} Mattiussi C {1997} \tia{An analysis of finite volume, finite element, and finite difference methods using some concepts from algebraic topology}
\doin{10.1006/jcph.1997.5656}{J.\ Comput.\ Phys.}{}{133}{289}
\bibitem{Teixeira:1999hv} Teixiera F L and Chew W C {1999} \tia{Lattice electromagnetic theory from a topological viewpoint}  \doin{10.1063/1.532767}{J.\ Math.\ Phys.}{}{40}{169}
\bibitem{Bombelli:2009hg} Bombelli L, Corichi A and Winkler O {2009} \tia{Semiclassical quantum gravity: obtaining manifolds from graphs}
  \doin{10.1088/0264-9381/26/24/245012}{Class.\ Quantum Grav.}{}{26}{025012} (\arX{0905.3492})
\bibitem{Chung:1997tk} Chung F R K 1997 \emph{Spectral Graph Theory} (Providence, RI: American Mathematical Society)
\bibitem{Wardetzky:2008kk} Wardetzky M, Mathur S, K{\"a}lberer F and Grinspun E 2007 \tia{Discrete
  Laplace operators: no free lunch} \emph{Eurographics Symposium on Geometry Processing} ed A Belyaev and M Garland (New York: ACM)
\bibitem{Kigami:2001wk} Kigami J, 2001 \emph{Analysis on Fractals} (Cambridge: Cambridge University Press)
\bibitem{Osterwalder:1973tq} Osterwalder K and Schrader R {1973} \tia{Axioms for Euclidean Green's functions}
   \doin{10.1007/BF01645738}{Commun.\ Math.\ Phys.}{}{31}{83}
\bibitem{CES} Calcagni G, Eichhorn A and Saueressig F 2013 Probing the quantum nature of spacetime by diffusion \arX{1304.7247}
\bibitem{frc1} Calcagni G {2012} \tia{Geometry of fractional spaces}
\ndoin{http://intlpress.com/site/pub/pages/journals/items/atmp/content/vols/0016/0002/00024226/index.html}{Adv.\ Theor.\ Math.\ Phys.}{}{16}{549} (\arX{1106.5787})
 \bibitem{frc2} Calcagni G 2012 \tia{Geometry and field theory in multi-fractional spacetime}
 \doij{10.1007/JHEP01(2012)065}{J.\ High Energy Phys.}{}{JHEP01}{065}{2012} (\arX{1107.5041})
\bibitem{fra6}  Calcagni G {2012} \tia{Diffusion in quantum geometry} \doin{10.1103/PhysRevD.86.044021}{Phys.\ Rev.}{D}{86}{044021} (\arX{1204.2550})
\bibitem{frc3}  Calcagni G and Nardelli G 2012 \tia{Momentum transforms and Laplacians in fractional spaces} \ndoin{http://intlpress.com/site/pub/pages/journals/items/atmp/content/vols/0016/0004/00026468/index.html}{Adv.\ Theor.\ Math.\ Phys.}{}{16}{1315} (\arX{1202.5383})
\bibitem{GuR} Gurau R and Ryan J P 2013 \tia{Melons are branched polymers} \arX{1302.4386}
\bibitem{DJW1}  Durhuus B, Jonsson T and Wheater J F {2006} \tia{Random walks on combs} \doin{10.1088/0305-4470/39/5/002}{J.\ Phys.\ A: Math.\ Gen.}{}{39}{1009} (\arX{hep-th/0509191})
\bibitem{AGW}   Atkin M R, Giasemidis G and Wheater J F {2011} \tia{Continuum random combs and scale dependent spectral dimension} \doin{10.1088/1751-8113/44/26/265001}{J.\ Phys.\ A: Math.\ Theor.}{}{44}{265001} (\arX{1101.4174})
\bibitem{GWZ1}  Giasemidis G, Wheater J F and Zohren S {2012} \tia{Dynamical dimensional reduction in toy models of $4D$ causal quantum gravity} \doin{10.1103/PhysRevD.86.081503}{Phys.\ Rev.}{D}{86}{081503} (\arX{1202.2710})
\bibitem{GWZ2}  Giasemidis G, Wheater J F and Zohren S {2012} \tia{Multigraph models for causal quantum gravity and scale dependent spectral dimension} \doin{10.1088/1751-8113/45/35/355001}{J.\ Phys.\ A: Math.\ Theor.}{}{45}{355001} (\arX{1202.6322})
\bibitem{Regge:1961ct} Regge T {1961} \tia{General relativity without coordinates}
\doin{10.1007/BF02733251}{Nuovo Cimento}{}{19}{558}
\bibitem{Loll:1998ue} Loll R {1998} \tia{Discrete approaches to quantum gravity in four dimensions}
  \doin{10.12942/lrr-1998-13}{Living Rev.\ Rel.}{}{1}{13} 
\bibitem{Sorkin:1975kv} Sorkin R {1975} \tia{The electromagnetic field on a simplicial net}
\doin{10.1063/1.522483}{J.\ Math.\ Phys.}{}{16}{2432}
\bibitem{Dittrich:2012uj} Dittrich B {2012} \tia{From the discrete to the continuous ---towards a
  cylindrically consistent dynamics} \doin{10.1088/1367-2630/14/12/123004}{New J.\ Phys.}{}{14}{123004} (\arX{1205.6127})
\bibitem{haus} Steinhaus S 2012 private communication
\bibitem{Ambjorn:2011wg} Ambj{\o}rn J, Jurkiewicz J and Loll R {2011} \tia{Lattice quantum gravity ---an
  update}  \ndoin{http://pos.sissa.it/cgi-bin/reader/conf.cgi?confid=105}{PoS}{}{(Lattice 2010)}{014} (\arX{1105.5582})
\bibitem{Caselle:1989cd} Caselle M, D'Adda A and Magnea L {1989} \tia{Regge calculus as a local theory of
  the Poincar{\'e} group} \doin{10.1016/0370-2693(89)90441-3}{Phys.\ Lett.}{B}{232}{457}
\bibitem{Barrett:1999ba} Barrett J W {1999} \tia{First order Regge calculus}
\doin{10.1088/0264-9381/11/11/013}{Class.\ Quantum Grav.}{}{11}{112723} (\arX{hep-th/9404124})
\bibitem{Gionti:2005gi} Gionti G {2005} \tia{Discrete gravity as a local theory of the Poincar{\'e} group
  in the first-order formalism} \doin{10.1088/0264-9381/22/20/004}{Class.\ Quantum Grav.}{}{22}{204217}
  (\arX{gr-qc/0501082})
\bibitem{Baratin:2010ti} Baratin A and Oriti D {2010} \tia{Group field theory with non-commutative metric variables}
  \doin{10.1103/PhysRevLett.105.221302}{Phys.\ Rev.\ Lett.}{}{105}{221302} (\arX{1002.4723})
\bibitem{Baratin:2011hc} Baratin A, Dittrich B, Oriti D and Tambornino J {2011} \tia{Non-commutative
  flux representation for loop quantum gravity}
  \doin{10.1088/0264-9381/28/17/175011}{Class.\ Quantum Grav.}{}{28}{5011} (\arX{1004.3450})
\bibitem{Barrett:1998fp} Barrett J W and Crane L {1998} \tia{Relativistic spin networks and quantum gravity}
  \doin{10.1063/1.532254}{J.\ Math.\ Phys.}{}{39}{3296} (\arX{gr-qc/9709028})
\bibitem{Barrett:1999fa} Barrett J W, Rocek M and Williams R M {1999} \tia{A note on area variables in
  Regge calculus} \doin{10.1088/0264-9381/16/4/025}{Class.\ Quantum Grav.}{}{16}{1373}
  (\arX{gr-qc/9710056})
\bibitem{Makela:1994hm} M{\"a}kel{\"a} J {1994} \tia{Phase space coordinates and the Hamiltonian constraint
  of Regge calculus}  \doin{10.1103/PhysRevD.49.2882}{Phys.\ Rev.}{D}{49}{2882}
\bibitem{Dittrich:2008hg} Dittrich B and Speziale S {2008} \tia{Area-angle variables for general
  relativity}  \doin{10.1088/1367-2630/10/8/083006}{New\ J.\ Phys.}{}{10}{3006}
  (\arX{0802.0864})
\bibitem{Bianchi:2008ib} Bianchi E {2008} \tia{The length operator in loop quantum gravity}
\doin{10.1016/j.nuclphysb.2008.08.013}{Nucl.\ Phys.}{B}{807}{591} (\arX{0806.4710})
\bibitem{Tikhonov:1977ug} Tikhonov A N and Arsenin V I A 1977 \emph{Solutions of Ill-Posed Problems} (New York: Wiley)
\bibitem{Oriti:2012kx} Oriti D, Pereira R and Sindoni L {2012} \tia{Coherent states in quantum gravity:
  a construction based on the flux representation of loop quantum gravity}
  \doin{10.1088/1751-8113/45/24/244004}{J.\ Phys.\ A: Math.\ Theor.}{}{45}{4004} (\arX{1110.5885})
\bibitem{Oriti:2012hl} Oriti D, Pereira R and Sindoni L {2012} \tia{Coherent states for quantum
  gravity: toward collective variables}
  \doin{10.1088/0264-9381/29/13/135002}{Class.\ Quantum Grav.}{}{29}{135002} (\arX{1202.0526})
\bibitem{Dittrich:2008pw} Dittrich B {2009} \tia{Diffeomorphism symmetry in quantum gravity models}
\doin{10.1166/asl.2009.1022}{Adv.\ Sci.\ Lett.}{}{2}{151} (\arX{0810.3594})
\bibitem{Dupuis:2012ub} Dupuis M, Ryan J P and Speziale S {2012} \tia{Discrete gravity models and loop
  quantum gravity: a short review} \doin{10.3842/SIGMA.2012.052}{SIGMA}{}{8}{052} (\arX{1204.5394})
\bibitem{Ponzano:1968wi} Ponzano G and Regge T 1968 \tia{Semiclassical limit of Racah coefficients} \emph{Spectroscopic and Group Theoretical Methods in Physics} ed F Bloch (New York: Wiley). 
\bibitem{Noui:2005js} Noui K and P{\'e}rez A {2005} \tia{Three-dimensional loop quantum gravity:
  physical scalar product and spin-foam models}
  \doin{10.1088/0264-9381/22/9/017}{Class.\ Quantum Grav.}{}{22}{1739} (\arX{gr-qc/0402110})
\bibitem{Strichartz:2006tm} Strichartz R S 2006 \emph{Differential Equations on Fractals} (Princeton, NJ: Princeton University Press)
\bibitem{Hamber:2007wi} Hamber H W 2007 \tia{Discrete and continuum quantum gravity}
 \arX{0704.2895}
\bibitem{AltshillerCourt:1964vd} Altshiller-Court N 1964 \emph{Modern Pure Solid Geometry} (New York: Chelsea)
\end{thebibliography}
\end{document}